\documentclass[twocolumn,english,prl,floatfix,twocolumn,showpacs]{revtex4-1}
\usepackage[T1]{fontenc}
\usepackage[latin9]{inputenc}
\usepackage{amsbsy}
\usepackage{amstext}
\usepackage{amsmath}
\usepackage{graphicx}
\usepackage{color}
\makeatletter

\@ifundefined{textcolor}{}
{%
 \definecolor{BLACK}{gray}{0}
 \definecolor{WHITE}{gray}{1}
 \definecolor{RED}{rgb}{1,0,0}
 \definecolor{GREEN}{rgb}{0,1,0}
 \definecolor{BLUE}{rgb}{0,0,1}
 \definecolor{CYAN}{cmyk}{1,0,0,0}
 \definecolor{MAGENTA}{cmyk}{0,1,0,0}
 \definecolor{YELLOW}{cmyk}{0,0,1,0}
}

\newcommand{\be}{\begin{eqnarray}}
\newcommand{\ee}{\end{eqnarray}}\def\beq{\begin{equation}}\def\eeq{\end{equation}}

\usepackage{babel}
\makeatother

\begin{document}

\title{Two-channel Kondo physics due to As vacancies in the layered compound   ZrAs$_{1.58}$Se$_{0.39}$}

\author{T.\,Cichorek}
\affiliation{Institute of Low Temperature and Structure Research, Polish Academy of  Sciences, 50-950 Wroc\l{}aw, Poland}
\author{L.\,Bochenek}
\affiliation{Institute of Low Temperature and Structure Research, Polish Academy of  Sciences, 50-950 Wroc\l{}aw, Poland}

\author{M.\,Schmidt}\affiliation{Max-Planck-Institute for Chemical Physics of Solids, 01187 Dresden, Germany}

\author{R.\,Niewa}\affiliation{Institute of Inorganic Chemistry, University of Stuttgart, 70569 Stuttgart, Germany}

\author{A.\,Czulucki}\affiliation{Max-Planck-Institute for Chemical Physics of Solids, 01187 Dresden, Germany}

\author{G.\,Auffermann}\affiliation{Max-Planck-Institute for Chemical Physics of Solids, 01187 Dresden, Germany}

\author{F.\,Steglich}\affiliation{Max-Planck-Institute for Chemical Physics of Solids, 01187 Dresden, Germany}
\affiliation{Center for Correlated Matter, Zhejiang University, Hangzhou, Zhejiang 310058, China}
\affiliation{Institute of Physics, Chinese Academy of Science, Beijing 100190, China}

\author{R.\,Kniep}
\affiliation{Max-Planck-Institute for Chemical Physics of Solids, 01187 Dresden, Germany}

\author{S.\,Kirchner}\email{stefan.kirchner@correlated-matter.com}
\affiliation{Center for Correlated Matter, Zhejiang University, Hangzhou, Zhejiang 310058, China}

\date{\today}

\begin{abstract}
We address the origin of the magnetic-field independent $-|A|\,T^{1/2}$ term observed in the low-temperature resistivity of several As-based metallic systems of the PbFCl structure type. 
For the layered compound ZrAs$_{1.58}$Se$_{0.39}$, we show that vacancies in the square nets of As give rise to the low-temperature transport anomaly over a wide temperature regime of almost two decades in temperature. This low-temperature behavior is in line with the non-magnetic version of the two-channel Kondo effect, whose origin we ascribe to a dynamic Jahn-Teller effect operating at the vacancy-carrying As layer with a $C_4$ symmetry. The pair-breaking nature of the dynamical defects in the square nets of As explains the 
 low superconducting transition temperature $T_{\rm{c}}\approx 0.14$\,K of  ZrAs$_{1.58}$Se$_{0.39}$, as compared to the free-of-vacancies homologue ZrP$_{1.54}$S$_{0.46}$ ($T_{\rm{c}}\approx 3.7$\,K). Our findings should be relevant to  a wide class of metals with disordered pnictogen layers.

\end{abstract}


\maketitle

In the last years, several exciting phenomena have been discovered in pnictogen-containing materials. This holds particularly true for high-temperature superconductivity in a class of materials based on iron \cite{Kamihara.08}.  Another example is the family of filled skutterudites with pnictogen atoms in the cage~\cite{Sato,Shi.11}.
Topologically nontrivial phases of certain 3D insulators as well as Dirac and Weyl semimetals have been first observed in pnictogen-based systems, such as Bi$_{1-x}$Sb$_x$ \cite{Hsieh.08}, Cd$_3$As$_2$  \cite{Liu.14}, and TaAs  \cite{Xu.15}. 
Here, we address the origin of the magnetic-field independent $-|A|\,T^{1/2}$ term observed in the low-temperature resistivity of several As-based metallic systems. We show that As vacancies in the layered compound ZrAs$_{1.58}$Se$_{0.39}$ give rise to an orbital  two-channel Kondo effect (2CK) that is symmetry-protected against level splitting \cite{Aleiner2002,Moustakas96}

The standard theory of metals states that
at sufficiently low temperatures ($T$),
the electrical resistivity  $\rho(T)$ is expected to vary quadratically with $T$ due to the electron-electron interaction (EEI)~\cite{Hewson}. 
Deviations from this behavior can occur, but generally
 $\rho(T)$ continues to decrease upon cooling. 
Only a few mechanisms are known to produce a low-$T$ resistivity minimum. The paradigmatic example is the Kondo effect where spin-flip scattering of conduction electrons off dynamic centers associated with the local magnetic moments gives rise to a logarithmic increase of the resistivity upon cooling~\cite{Hewson}. Theoretical studies have revealed that an even more exotic 2CK effect can occur if two degenerate channels of the conduction electrons exist, which independently scatter off centers with a local quantum degrees of freedom \cite{Nozieres.80}.
The low-energy physics in this case is governed by a non-Fermi liquid fixed point and results in a resistivity minimum followed at lower temperature by a $\sqrt{T}$ increase of $\rho(T)$. 
Realizing the nonmagnetic version of the 2CK effect requires a degeneracy between orbitals that is difficult to maintain in real systems.
Here, the orbital degree
of freedom plays the role of the (pseudo-)spin and the  electron spin  represents the degenerate channel index. Such a scenario was proposed for metals with dynamic structural defects modeled by double-well potentials~\cite{Zawadowski1980,Vladar1983a,Vladar1983b}. However, in spite of considerable interest~\cite{Cochrane1975,Ralph1994,Halbritter.00,Huang2007,Zhu.16},  this type of 2CK physics has never been conclusively demonstrated in any bulk metallic system.  This appears to be in line with further theoretical efforts that established  two-level systems as an unlikely source of the non-Fermi behavior~\cite{Moustakas96,Aleiner2001,Aleiner2002,Zarand2005}. Additional complications arise from the fact that the disorder-enhanced EEI in three-dimensional metals can also give rise to a $\sqrt{T}$ term in the resistivity~\cite{Altshuler1985,Lee1985}. 
Thus, the clearest evidence to date for the 2CK effect has come from an artificial nanostructure that can be judiciously tuned to show a $T^{-1/2}$ term in the differential conductance for about a decade in $T$~\cite{Potok.06}.\\
On the other hand, there are a number of experimental results 
which point to  unconventional scattering mechanisms whose precise nature has remained enigmatic~\cite{Lin2002}. In particular, several As-based metallic systems with the layered PbFCl
structure display a low-\textit{T} resistivity minimum.
Investigations on e.g. diamagnetic ThAsSe (nominal composition) single crystals revealed a magnetic-field-($B$)-independent $-|A|T^{1/2}$  term in the low-$T$ resistivity~\cite{Cichorek2005}. 
Furthermore, the thermal conductivity and specific heat of ThAsSe showed  glass-type temperature dependences which support the presence of structural defects with internal degrees of freedom \cite{Cichorek2005}. Investigations of the closely related Zr- and Hf-based arsenide selenides have pointed to a $B$-independent $-|A|T^{1/2}$  contribution to $\rho(T)$ as a generic feature of metallic arsenide selenides crystallizing in the PbFCl structure \cite{Czulucki2010}.
To address the physical mechanism leading to the $B$-independent $-|A|T^{1/2}$ term in the low-$T$ $\rho(T)$ we have identified two homologues,  ZrAs$_{1.58}$Se$_{0.39}$  and ZrP$_{1.54}$S$_{0.46}$, that allow us to identify the microscopic origin of the anomalous behavior. By combining precise physical property measurements with chemical and structural investigations performed on the same single crystals, we show that the only viable explanation for the observed transport anomalies  of ZrAs$_{1.58}$Se$_{0.39}$ is in terms of a 2CK.  We argue that 2CK physics is possible due to vacancies in the square nets of As atoms.

The ternary pnictide-chalcogenides ($Pn$--$Ch$) contain intermediate
phases which crystallize in the tetragonal
($P$4/$nmm$) PbFCl structure, a substitution
variant of the Fe$_2$As type. This crystal structure
consists of square-planar 4$^4$ nets stacked along the [001] direction.
For compounds with exact
chemical composition $M$:$Pn$:$Ch$=1:1:1, each of the
4$^4$ nets is exclusively occupied by one element resulting
in a layer-sequence ~...$Pn$--$M$--$Ch$--$Ch$--$M$--$Pn$, and forming a
puckered double-layer $M_2Ch_2$ with an ordered distribution
of $M$ and $Ch$.

Phases with $M$, being a metal with the fixed +4 oxidation state such as Zr, Hf, and Th, are characterized by significant excess of pnictogen atoms (see Fig.~1). In case that the homogeneity range is restricted to the tie-line between $MPn_2$ and $MCh_2$ the chemical formula corresponds to $MPn_xCh_y$ with $x + y = 2$. For $M$=Zr, it has been shown that only the 2$c$ site is randomly occupied by $Pn$ and $Ch$. Exceptionally for the arsenide selenide, however, the homogeneity range is located right next to the tie-line ZrAs$_2$--ZrSe$_2$ on the As depleted side and hence $1.90\leq x+y \leq 1.99$ ~\cite{Schlechte2007} and Fig.~S4 of \cite{SupMat}. As a result, vacancies are \textit{solely} present within the As layers (2$a$ site) \cite{SupMat}.

\begin{figure}[!ht]
\begin{center}
\includegraphics[width=\linewidth,keepaspectratio=true]{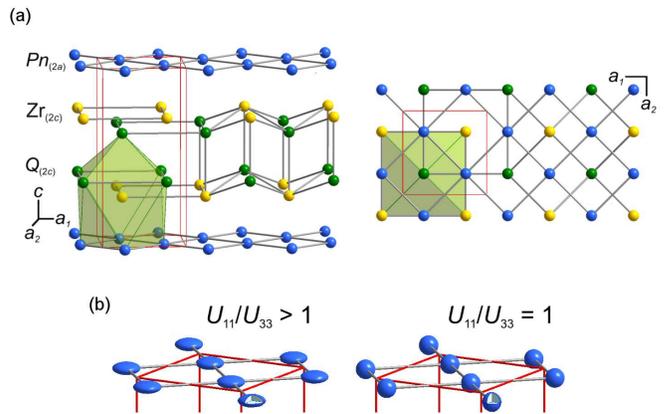}
\caption{Structural disorder in zirconium pnictide chalcogenides.
(a) The PbFCl structure type with the fully occupied \textit{Q}($2c$) site (green) by $Ch$  together with $Pn$. The \textit{Pn}($2a$) site (blue) arranged within planar layers is only occupied to 97\,\% by As in ZrAs$_{1.58}$Se$_{0.39}$, but 
fully occupied by P in ZrP$_{1.54}$S$_{0.46}$. 
(b) The vacancies in ZrAs$_{1.58}$Se$_{0.39}$ (left) manifest  random displacements of As within the layer due to homoatomar covalent bond formation. This is indicated by flattened displacement ellipsoids in the structure refinements, while the refined displacement ellipsoids in ZrP$_{1.54}$S$_{0.46}$ (right) are nearly spherical.}
\label{fig:Fig2}
\end{center}
\end{figure}

\begin{figure}
\begin{center}
\includegraphics[width=\linewidth,keepaspectratio=true]{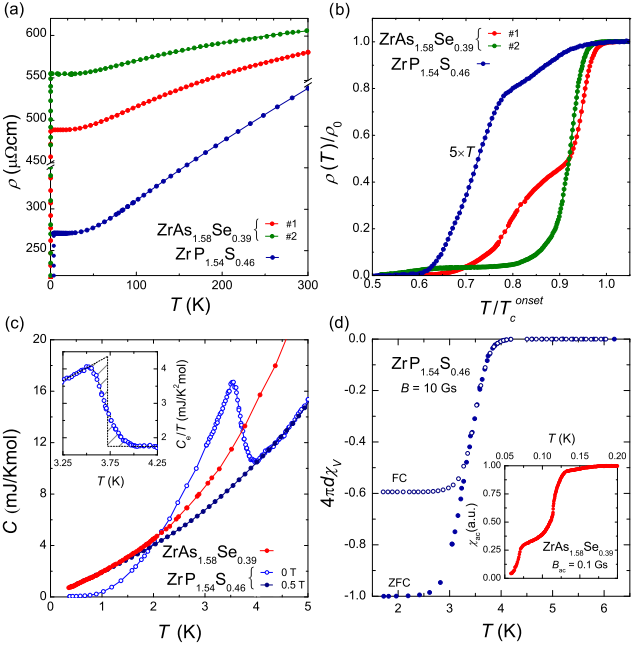}
\caption{Characteristics of the pnictide chalcogenide superconductors ZrAs$_{1.58}$Se$_{0.39}$ and ZrP$_{1.54}$S$_{0.46}$. 
(a) Normal-state electrical resistivity along the $c$ axis of ZrAs$_{1.58}$Se$_{0.39}$ and ZrP$_{1.54}$S$_{0.46}$ (the two specimens of the former material were cut off from the same single crystal). 
(b) The same results, but with focus on the vicinity of the superconducting transition, as $\rho/\rho_0$ vs. $T/T_{\rm{c}}^{
\rm{onset}}$. For clarity, the temperature scale of the $\rho(T)$ data for the P-based system was multiplied by a factor of  
5.  
(c) Low-temperature specific heat  for ZrAs$_{1.58}$Se$_{0.39}$ and ZrP$_{1.54}$S$_{0.46}$. Inset: Electronic specific heat of ZrP$_{1.54}$S$_{0.46}$ in the vicinity of the superconducting phase transition, as $C_{\rm{e}}/T$ vs temperature. $C_{\rm{e}}=C-C_{\rm{ph}}$, where $C_{\rm{ph}}$ is the phonon contribution, estimated from the normal-state $B=0.5$\,T data.
(d)  Zero-field-cooled (ZFC) and field-cooled (FC)  dc magnetic susceptibility as a function of temperature for ZrP$_{1.54}$S$_{0.46}$.  Inset: Evidence for large shielding in ZrAs$_{1.58}$Se$_{0.39}$  is provided by the sizable diamagnetic signal of the ac magnetic susceptibility below about $T = 0.125$\,K. 
}
\label{fig:Fig3}
\end{center}
\end{figure}

\begin{figure}[!ht]
\begin{center}
\includegraphics[width=\linewidth,keepaspectratio=true]{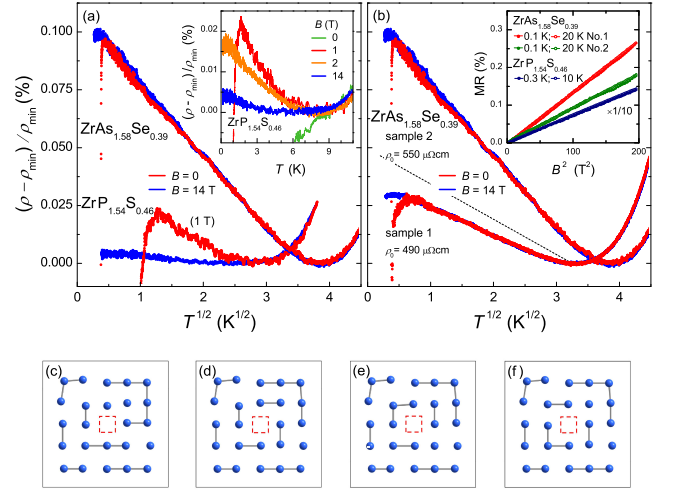}
\caption{Evidence for the 2CK effect due to As vacancies. 
(a) Despite very similar basic low-temperature properties of ZrAs$_{1.58}$Se$_{0.39}$ and ZrP$_{1.54}$S$_{0.46}$, these sister compounds display qualitatively different responses of $(\rho-\rho_{\rm{min}})/\rho_{\rm{min}}$ vs. $T^{1/2}$. Note the magnetic-field-independent 
$-|A|T^{1/2}$ contribution to $\rho(T)$ for ZrAs$_{1.58}$Se$_{0.39}$, i.e., for a system with vacancies in the pnictogen layer. Inset: Low-$T$ resistivity of ZrP$_{1.54}$S$_{0.46}$ measured in zero and varying magnetic
fields up to 14\,T. For $B = 0$, $\rho_{\rm{min}}=\rho$(9\,K) was taken.
(b) Remarkable differences of the $B$-independent low-$T$ $\rho(T)$ as $(\rho-\rho_{\rm{min}})/\rho_{\rm{min}}$ vs. $T^{1/2}$ for two single-crystalline ZrAs$_{1.58}$Se$_{0.39}$  specimens with similar residual resistivities. Dashed line shows the magnitude of a hypothetical $-|A|T^{1/2}$ correction to $\rho(T)$ for sample \#2 due to 3D EEI assuming $\tilde{F}_\sigma=0$. Inset: Magnetic field dependence of $\rho(T)$ of ZrAs$_{1.58}$Se$_{0.39}$ and ZrP$_{1.54}$S$_{0.46}$ single crystals. 
Note that the data for ZrAs$_{1.58}$Se$_{0.39}$  are divided by a factor of 10.
(c)-(f) Dynamic structural scattering centers in metallic arsenide selenides. Shown are different possible arrangements of dimers or oligomers within the As layer triggered by the  vacancy (red square). Arsenic atoms may for example arrange as dimers, trimers, oligomers of several As atoms, or infinite chains with zigzag or sawtooth configuration. (f) is obtained from (e) by a $\pi/2$ rotation of the arrangement of (e) indicating how the $C_4$ symmetry is dynamically restored.}
\label{fig:Fig4}
\end{center}
\end{figure}

Replacement of As and Se by P and S, respectively, does not distinctly alter basic physical properties, such as electrical resistivity and specific heat of the tetragonal Zr--$Pn$--$Ch$ phases. In fact, differences in the metallic behavior of $\rho(T)$  are of minor significance only, as depicted in Fig.~\ref{fig:Fig3}(a). 
For both homologues  Zr$Pn_xCh_y$, a drop of $\rho(T)$ signals the onset of a superconducting phase transition at $T_{\rm{c}}\approx 0.14$\,K (ZrAs$_{1.58}$Se$_{0.39}$) and 3.9\,K (ZrP$_{1.54}$S$_{0.46}$), respectively. 
However, $\rho=0$ is found only substantially below the onset temperature [see Fig.~\ref{fig:Fig3}(b)]. The complex behavior of $\rho(T)$  in the vicinity of $T_{\rm{c}}$ is ascribed to a likely delicate variation of chemical composition,  i.e., below 0.5 wt.\,\% which is the resolution limit of our analysis. 
The normal-state specific heats $C(T)$ for both ZrAs$_{1.58}$Se$_{0.39}$ and ZrP$_{1.54}$S$_{0.46}$ are very similar, leading to virtually the same value of the Sommerfeld coefficient of the electronic specific heat $\gamma  = 1.7 (\pm 0.1)$\,mJK$^{-2}$mol$^{-1}$ [see Fig.~\ref{fig:Fig3}(c)]. 

Summing up, the main physical properties of ZrAs$_{1.58}$Se$_{0.39}$ and ZrP$_{1.54}$S$_{0.46}$ demonstrate a far-reaching similarity between both homologues. Therefore, the observed difference in their $T_{\rm{c}}$'s by a factor $\approx 30$ is an unexpected result, especially in view of the tiny concentration of magnetic impurities in the arsenide selenide crystals, i.e. less than 0.10 wt.\,\%~\cite{Czulucki2010}). Note that  the As vacancies themselves, as (static) potential scatterers, cannot be the source of strong pair-breaking.

A hallmark of the 2CK effect is the $\sqrt{T}$ dependence of the low-$T$ resistivity. In the case of  dynamic structural defects, the resulting non-Fermi liquid properties are not expected to depend  on an applied magnetic field as long as the Zeeman splitting does not cause a difference in the conduction-electron DOS at the Fermi level between up and down subbands. The only other mechanism that can in principle result in a $B$-independent $-|A|T^{1/2}$ correction to the resistivity
is the  electron-electron interaction (EEI) in a three-dimensional, disordered metal \cite{Altshuler1985,Lee1985}.
This could happen if electron screening is strongly reduced, yielding the unique case of a screening factor, $\tilde F_{\sigma}$, very close  or even equal to zero~\cite{Lee1985,Altshuler1985}. To analyze possible corrections due to the EEI in ZrAs$_{1.58}$Se$_{0.39}$ and ZrP$_{1.54}$S$_{0.46}$, we thus have plotted in Fig.~\ref{fig:Fig4}(a)
the relative change of the resistivity normalized to the minimum value, $(\rho-\rho_{\rm{min}})/\rho_{\rm{min}}$, as a function of $T^{1/2}$. 
Below $T_{\rm{min}}\approx 15.0$\,K, the zero-field $\rho(T)$ data for ZrAs$_{1.58}$Se$_{0.39}$ depend strictly linearly on $T^{1/2}$ over almost two decades in temperature. A magnetic field of $B = 14$\,T, being the largest field accessible in our experiment, does not alter the 
$-|A|T^{1/2}$  dependence for the system containing vacancies in the pnictogen layers. 

For the free-of-vacancies system, the low-$T$ upturn  displays a completely different response to an applied field. (In the absence of a magnetic field, superconducting fluctuations dominate the transport properties of ZrP$_{1.54}$S$_{0.46}$  leading to a strong decrease of $\rho(T)$ in a temperature region which significantly exceeds $T_{\rm{c}}\approx 3.7$\,K.) Although at $B = 1$\,T a small fraction of the ZrP$_{1.54}$S$_{0.46}$  sample still displays surface superconductivity below 1\,K,
at elevated temperatures, we observe an upturn in $\rho(T)$ with a rather complex $T$ behavior. More remarkably, however, the $(\rho-\rho_{\rm{min}})/\rho_{\rm{min}}$ correction is distinctly smaller at intermediate fields of about 2\,T [cf. the inset of Fig. 3(a)]. For 14\,T, the low-$T$ upturn is strongly reduced, and its magnitude is about 7 times smaller than for $B = 1$\, T. Furthermore, the  $\rho(T)$ rise is restricted to a rather narrow temperature window, i.e. from above 4.2\,K to around 1\,K. At $T< 1$\,K, the resistivity tends to saturate.
These findings point to a negligible EEI in single crystals of Zr$Pn_xCh_y$ and yield striking evidence for an entirely different origin of the $-|A| T^{1/2}$ term in the low-$T$ $\rho(T)$  occurring in the material without (ZrAs$_{1.58}$Se$_{0.39}$) and with (ZrP$_{1.54}$S$_{0.46}$) a full occupancy of the square-planar pnictogen layers. For an in-depth analysis, see \cite{SupMat}. In the latter case, structural disorder (driven from the mixed occupied 2$c$ sites only) leads to weak localization. Its  negative contribution to the magnetoresistance MR$=[(\rho(B)-\rho(0)]/\rho(0)$ increases upon cooling and amounts to about 0.02\,\%  at $T\,=\,2$\, K. The weak-localization contribution is  nearly two orders of magnitude  smaller than the classical MR [cf. the inset of Fig.~3(b)]. Therefore, the total MR of  ZrP$_{1.54}$S$_{0.46}$  is essentially temperature independent at $T\leq 10$\, K and approaches 1.4\,\% at $B = 14$\,T. 

Figure~3(b) shows the temperature dependence of $(\rho-\rho_{\rm{min}})/\rho_{\rm{min}}$  for two ZrAs$_{1.58}$Se$_{0.39}$ specimens with similar elastic relaxation times. 
In spite of this moderate variation, the size of the magnetic-field-insensitive $-|A|T^{1/2}$ term between these samples differs by more than a factor of 3 and hence the $A$ coefficient amounts to 0.038 and $0.167~\Omega$cm/K$^{1/2}$ for samples \#1 and \#2, respectively. Similarly to the afore mentioned results, the experimental observations shown in Fig.~3(b) are at strong variance to the expectation based on enhanced EEI in 3D specimens of the same disordered metal \cite{Altshuler1985,Lee1985}. Indeed, a nearly identical screening factor $\tilde F_{\sigma}$, see Ref.~\cite{SupMat}, implies that the magnitude of a hypothetical $(\rho-\rho_{\rm{min}})/\rho_{\rm{min}}$ anomaly would only vary with  $\rho D^{-1/2} \propto \rho^{3/2}$. This, however, is not observed in ZrAs$_{1.58}$Se$_{0.39}$. In fact, a supposed 
$-|A|T^{1/2}$ correction, calculated in respect to the $|A|$-coefficient value of sample \#1 and schematically sketched by a dashed line in Fig.~3(b), would be substantially smaller than what is experimentally found for sample \#2. (Since $A$ is field independent, $\tilde F_{\sigma} = 0$ was assumed in our calculations \cite{SupMat}).

We thus conclude that the $B$-field independent $-|A|T^{1/2}$ correction to $\rho(T)$ in ZrAs$_{1.58}$Se$_{0.39}$ cannot be caused by EEI and can only be explained by the existence of non-magnetic defects with degenerated ground state which, at low temperatures, place the system near the 2CK fixed point. Note that in the dilute limit, where the dynamic scattering centers are independent of each other, the amplitude $|A|$ is proportional to the concentration of dynamic scattering centers in the 2CK regime which is in general smaller than, and not expected to scale in a simple fashion with, the concentration of As vacancies.

The chemical composition of ZrAs$_{1.58}$Se$_{0.39}$ implies that the vacancies in the As layer are in the dilute limit. As discussed above, our analysis shows that interstitial As does not occur in the pnictogen layer.
 Each vacancy thus preserves the $C_4$ symmetry of the pnictogen layer. As a result, a Jahn-Teller distortion forms in the pnictogen layer concomitant with a formation of As dimers or oligomers. This phenomenon is well known to occur in such square nets of the PbFCl structure-type compounds~\cite{Tremel1987}. The flattened displacement ellipsoids shown in Fig.~1(b)  are indicative of the occurrence of the  dynamic Jahn-Teller effect in the As (2$a$) layer. As the Jahn-Teller distortion develops, the doublet states split and one of them becomes the new ground state. A finite tunneling rate between the different impurity positions compatible with the overall square symmetry restores the square symmetry through a dynamic Jahn-Teller effect, see Figs.~\ref{fig:Fig4}(c)-(f). The group $C_4$ possesses only one-dimensional irreducible representations and one two-dimensional irreducible representation (IRREP).
Thus, a non-Kramers doublet associated with the dynamic Jahn-Teller distortion transforms as the two-dimensional irreducible representation of the group $C_4$, and allows for a two-channel Kondo fixed point to occur at sufficiently low energies~\cite{Gogolin.96,Moustakas.97,Hotta2006}. The pseudo-spin index $\pm$ of the effective 2CK Hamiltonian labels the basis states of this two-dimensional subspace~\cite{SupMat}. As a result of the coupling to the conduction electrons the doublet can become the ground state~\cite{Arnold2007}. In fact, the doublet is renormalized below the singlet in a wide parameter regime~\cite{FuhChio.14}.

Our conclusion that 2CK centers exist in ZrAs$_{1.58}$Se$_{0.39}$ is further corroborated by the low superconducting transition temperature, as compared to ZrP$_{1.54}$S$_{0.46}$, which points to the presence of  efficient Cooper pair breakers \cite{Sellier2001}.
The 2CK effect in ZrAs$_{1.58}$Se$_{0.39}$ has been argued to arise out of the non-Kramers doublet transforming as the two-dimensional IRREP of $C_4$. The associated basis states of the two-dimensional IRREP, labeled by $+$ and $-$, are time-reversed partners. 
Expanding the BCS  order parameter around the quantum defect will thus have to involve singlets of $+$ and $-$. The full Hamiltonian including the quantum defect involves scattering from one of the basis states of the two-dimensional IRREP to the other. These scattering processes therefore have to break up Cooper pairs  
and thus reduces $T_{\rm{c}}$. For details, see \cite{SupMat}.
Thus, our model is in line with all observed properties of ZrAs$_{1.58}$Se$_{0.39}$ and is not susceptible to the difficulties  that exist for the 2CK scenario based on two-level systems. The observed suppression of superconductivity in ZrAs$_{1.58}$Se$_{0.39}$  as compared to ZrP$_{1.54}$S$_{0.46}$ is naturally explained.
A log$T$-behavior of $\rho(T)$ is expected for the $T$-behavior right above the $\sqrt{T}$ regime for a 2CK system, which is indeed observed \cite{SupMat}.
Direct evidence of the dynamic scattering centers could come from scanning tunneling spectroscopy apt  to zoom into one of these dynamic scattering centers.

In conclusion, we have shown that the $B$-independent $-|A|T^{1/2}$-term in $\rho(T)$ of ZrAs$_{1.58}$Se$_{0.39}$ is triggered by non-magnetic centers with a local quantum degree  of freedom. Our analysis indicates that a dynamic Jahn-Teller effect concomitant with the formation of As oligomers places the scattering center in the vicinity of the orbital two-channel Kondo fixed point. These quantum impurities act as efficient Cooper pair breakers which explains the suppression of $T_c$ of ZrAs$_{1.58}$Se$_{0.39}$  as compared to the homologue  ZrP$_{1.54}$S$_{0.46}$
We expect that similar dynamic scattering centers  occur in other materials PbFCl structure type with square nets of pnictogen.

We greatly acknowledge helpful discussions with Z.\,Henkie, J.\,Kroha, J.\,J.\,Lin, P.\, Ribeiro, F.\ Zamani and A.\,Zawadowski. Experimental work on non-magnetic Kondo effect at the Institute of Low Temperature and Structure Research, Polish Academy of  Sciences in Wroclaw was supported by the Max Planck Society through the Partner Group Program.
S.\,Kirchner acknowledges partial support by the National Natural Science Foundation of China, grant No.11474250 and the National Science Foundation under Grant No. PHY11-25915.


\newpage
\clearpage

\begin{widetext}
\begin{center} {\large \bf Two-channel Kondo physics due to As vacancies in the layered compound   ZrAs$_{1.58}$Se$_{0.39}$\\ \large Supplementary Material }  

\bigskip 

T.\,Cichorek$^1$, L.\,Bochenek$^1$, M.\,Schmidt$^2$, R.\,Niewa$^3$, A.\,Czulucki$^2$, G.\,Auffermann$^2$, F.\,Steglich$^{2,4,5}$, R.\,Kniep$^2$ and S.~Kirchner$^{5}$\\
\medskip 
{\it \small
$^1$Institute of Low Temperature and Structure Research, Polish Academy of  Sciences, Wroc{\l}aw, Poland\\
$^2$Max Planck Institute for Chemical Physics of Solids, N\"othnitzer Stra\ss{}e 40, 01187 Dresden, Germany\\
$^3$Institute of Inorganic Chemistry, University of Stuttgart, Germany\\
$^4$Institute of Physics, Chinese Academy of Science, Beijing 100190, China\\
$^5$Center for Correlated Matter, Zhejiang University, Hanghzou, Zhejiang 310058, China
}
\end{center}

\setcounter{figure}{0}   \renewcommand{\thefigure}{S\arabic{figure}}
\setcounter{equation}{0} \renewcommand{\theequation}{S.\arabic{equation}}
\setcounter{section}{0} \renewcommand{\thesection}{S.\Roman{section}}
\renewcommand{\thesubsection}{S.\Roman{section}.\Alph{subsection}}
\makeatletter
\renewcommand*{\p@subsection}{}  
\makeatother
\renewcommand{\thesubsubsection}{S.\Roman{section}.\Alph{subsection}-\arabic{subsubsection}}
\makeatletter
\renewcommand*{\p@subsubsection}{}  
\makeatother
 \bibstyle{plainnat}
\renewcommand*{\citenumfont}[1]{S#1}
\renewcommand*{\bibnumfmt}[1]{[S#1]}

%
\section{I Crystal Growth, Chemical Characterization, and Structural Analysis}

Single crystals of the ternary phases with PbFCl structure type were grown by Chemical Transport Reaction (CTR) using iodine as transport agent~\cite{Binnewies} starting from microcrystalline powders of pre-reacted materials, which were obtained by gradual temperature treatment of mixtures of the elements under inert conditions (vacuum). The amount and chemical composition of deposited crystals depends on the quantity and composition of the source material. A large quantity of source material leads to almost constant transport conditions by keeping temperature of the source and sink as well as the total pressure in the transport ampoule constant. Various positions of the growing crystals within the sink area of the ampoule may cause differences in chemical composition due to different temperature gradients. For this reason and because of the homogeneity regions of the ternary phases under consideration, it is of crucial importance not only to check the total chemical composition of a crystal but to also investigate the constancy of the chemical composition over the crystal individual by WDXS-analyzes as a function of position. 
Microcrystalline samples of chemical composition ZrP$_{1.54}$S$_{0.46}$ and ZrAs$_{1.58}$Se$_{0.39}$ were prepared by reaction of the elements using glassy carbon crucibles as containers which were sealed in fused-silica ampoules. Starting from these powders, tetragonal single crystals (up to 2 mm in length) of the ternary phases were grown by exothermal CTR in temperature gradients from 875$^{\circ}\mathrm{C}$ (source) to 975$^{\circ}\mathrm{C}$ (sink) by using iodine as the transport agent.

 \begin{figure}[!ht]
\begin{center}
\includegraphics[width=0.5 \linewidth,keepaspectratio=true]{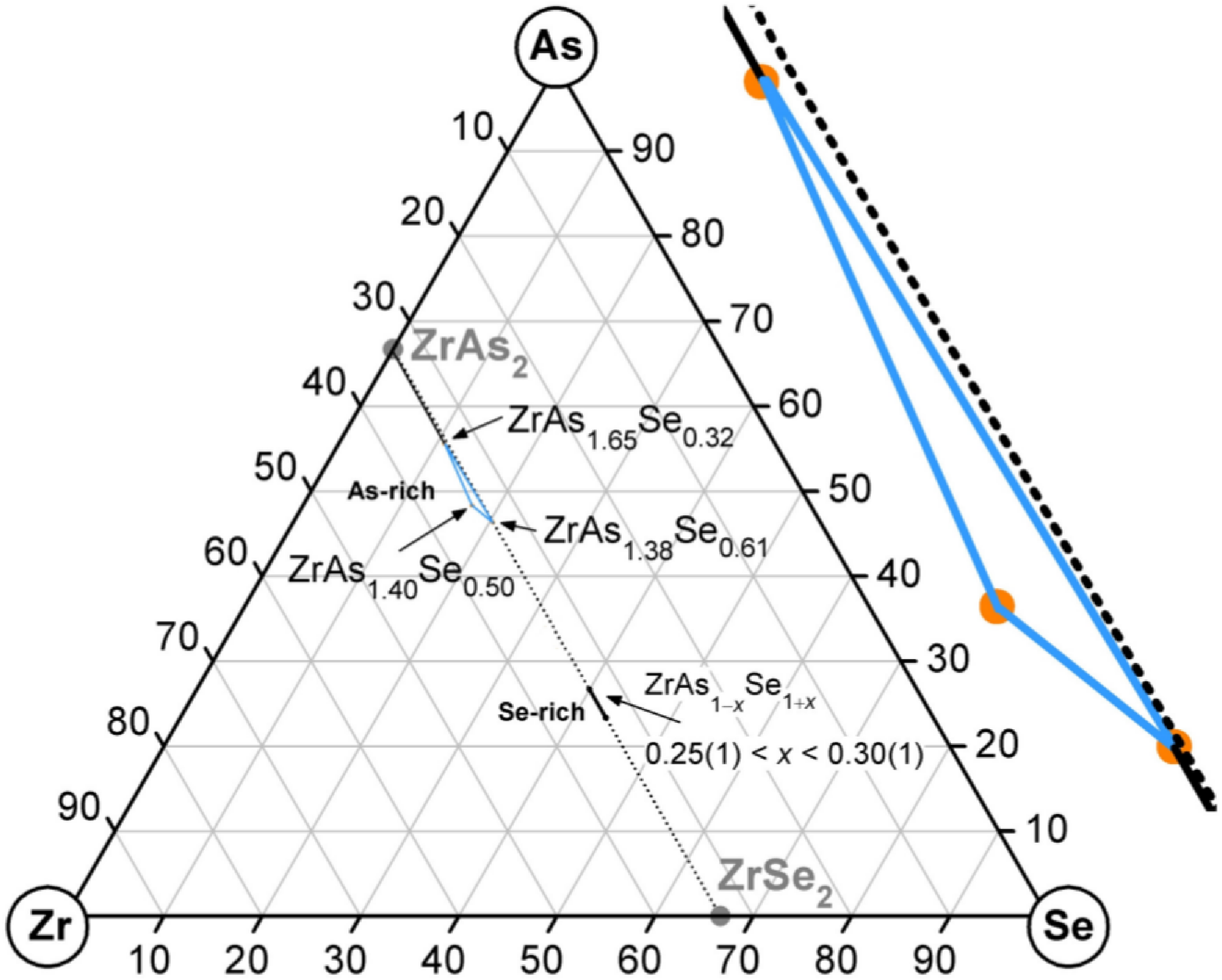}
\end{center}
\caption{
{\bf Isothermal section of the ternary system Zr--As--Se at 1223\,K.} The homogenity range of the PbFCl type ternary phase is
 enlarged, with the dashed line indicating the tie-line between the binary compounds ZrAs$_2$ and ZrSe$_2$~\cite{Schlechte2007}.
}
\label{fig:Fig1}
\end{figure}

Single crystals grown by the CTR in the system \mbox{Zr--P--S} reveal a rather close chemical composition with variations for Zr, P and S below 0.5\,wt\,\%: WDXS analyses (Cameca SX~100) on a single crystal resulted in Zr$_{1.000(4)}$P$_{1.543(5)}$S$_{0.460(4)}$; IC P-OES analyses (Vista RL, Varian) on crystals from the same batch gave Zr$_{1.00(2)}$P$_{1.56(1)}$S$_{0.461(4)}$, confirming an only narrow range in homogeneity. Crystallographic data: Tetragonal, $P$4/$nmm$, $a = 3.5851(1)$\,\AA, $c = 7.8036(2)$\,\AA, $Z = 2$.

A single crystal grown by the CTR in the system Zr--As--Se was investigated by WDXS. The chemical composition was determined as Zr$_{1.000(3)}$As$_{1.595(3)}$Se$_{0.393(1)}$ which is consistent with the results of crystal structure determination and refinements (ZrAs$_{1.59(1)}$ Se$_{0.39(1)}$, tetragonal, $P$4/$nmm$, $a = 3.7576(2)$\,\AA, $c = 8.0780(5)$\,\AA, $Z = 2$). 
A linescan of 50 measuring points over a distance of 500 $\mu$m along the crystal revealed a homogeneous distribution of Zr and Se, with only small fluctuations in the As content.


The ternary systems Zr--As--Se and Zr--P--S contain intermediate phases which crystallize in the tetragonal ($P$4/$nmm$) PbFCl~\cite{Wang.95} structure, a substitution variant of the Fe$_2$As type. The crystal structure consists of square-planar $4^4$ nets stacked along the [001] direction. For ternary pnictide-chalcogenides ($Pn$--$Ch$) with precise chemical composition $M$:$Pn$:$Ch$=1:1:1 each of the $4^4$ nets is exclusively occupied by one element resulting in a layer-sequence ...$Pn$--$M$--$Ch$--$Ch$--$M$--$Pn$, and forming a puckered double-layer $M_2Ch_2$ with an ordered distribution of $M$ and $Ch$. A representative for ternary pnictide-chalcogenides with precise 1:1:1 composition is given by CeAsSe~\cite{Schlechte2009}. 

The ternary phases adopting the PbFCl structure type (or more specifically the ZrSiS type) in the systems Zr--As--Se and Zr--P--S are characterized by homogeneity ranges and significant excess of $Pn$ (As and P, respectively). In case that the homogeneity range is restricted to the tie-line between Zr$Pn_2$ and Zr$Ch_2$ ($Ch$ = S and Se, respectively), the chemical formula of the ternary phase corresponds to Zr$Pn_xCh_y$ with $x + y = 2$, a situation which holds for the zirconium phosphide sulfide. The homogeneity range of the arsenide selenide, however, is shifted from the tie-line ZrAs$_2$--ZrSe$_2$ (see Fig.~1) and forms a triangle between the chemical compositions ZrAs$_{1.65}$Se$_{0.32}$, ZrAs$_{1.38}$Se$_{0.61}$ and ZrAs$_{1.40}$Se$_{0.50}$~\cite{Schlechte2007}. Detailed structural investigations on single crystals belonging to this homogeneity triangle revealed that the $Q$(2$c$) site (see Fig.~1 of the main text) is randomly (but fully) occupied by As and Se. At the same time, vacancies are present within the As layers (2$a$ site)~\cite{Schlechte2007}. In case of the ZrP$_x$S$_y$ ($x + y = 2$) series, vacancies within the P layers are missing (full occupation of the 2$a$ site by P).


Quadratic nets in metallic PbFCl-type compounds are expected to distort due to a second-order Jahn-Teller effect, i.e., chemical bond formation~\cite{Tremel1987} as was observed frequently in rare-earth metal chalcogenides~\cite{Doert2012} as well as CeAsSe~\cite{Schlechte2009}. However, for the compounds under consideration no crystallographic symmetry reduction, but rather local displacements of As atoms were observed in X-ray and neutron diffraction.
These displacements of As within the quadratic nets in the direction of neighboring As atoms indicate the presence of dimers or oligomers As$_n$, as illustrated in Figs.~3(d)-3(g) of the main text. The displacement factors of As within the plane of the 2$a$ sites clearly increase with increasing vacancy concentration, whereas the phase without a significant amount of vacancies,
namely Zr$_{1.000(4)}$P$_{1.543(5)}$S$_{0.460(4)}$, does not show any indication of P displacements.

\section{II Physical Measurements}

After the (destruction-free) chemical analysis and crystallographic investigations, measurements of physical properties were performed on one and the same crystal or by use of parts of the well-characterized specimen.

Specific heat  measurements 
in the range 0.4\,K\,$\leq$\,\textit{T}\,$\leq$\,5\,K were performed on a ZrAs$_{1.58}$Se$_{0.39}$ single crystal with a mass of 19.87(1) mg and a cubic-like shape of approximate edge length 1.4\,mm, and a single crystal of ZrP$_{1.54}$S$_{0.46}$ with a mass of 5.2(1)\,mg and a dimension of about 1.4\,mm along the \textit{c} axis via the thermal-relaxation method  using a commercial  $^{3}$He microcalorimeter\,(PPMS). 
For
ZrAs$_{1.58}$Se$_{0.39}$, the low-temperature ac susceptibility $\chi_{\rm{ac}}(T)$
was investigated with the driving field $B_{\rm{ac}}\,=\,0.01$\,mT utilizing a  $^{3}$He--$^{4}$He dilution refrigerator. 
For
ZrP$_{1.54}$S$_{0.46}$, dc magnetization measurements were performed for
1.8\,K$\leq$\,\textit{T}\,$\leq$\,6.5\,K and \textit{B}\,=\,1\,mT using a
superconducting quantum interference device magnetometer\,(MPMS). 

The electrical resistivity was studied by a standard
four-point ac technique in zero and applied magnetic fields up to 14\,T.
For low-temperature measurements, a Linear Research ac
resistance bridge (model 700) was utilized applying electrical currents as low as
150\,$\mu$A in the mK temperature range. In all samples, the resistivity
was measured parallel to the crystallographic \textit{c} axis. From the single crystal of
ZrAs$_{1.58}$Se$_{0.39}$ two specimens have been cut off, with a
length of 1.12\,mm\,(\#1) and 0.94\,mm\,(\#2) and a cross section of 0.023\,mm$^{2}$ and 0.032\,mm$^{2}$, respectively. For ZrP$_{1.54}$S$_{0.46}$, the specimen had a
length of 1.18\,mm  and a cross section of 0.15\,mm$^{2}$. Because of the high
fragility of the  crystals, electrical contacts for temperature-dependent
experiments were made by electrochemical deposition of copper, and a two-needle
voltage probe with a fixed distance of 0.548\,mm was used to determine absolute
values of the resistivity.

For the As-based system, the $c$-axis residual resistivity $\rho_0$ amounts to about $500\, \mu\Omega$\,cm. For the P-based system, one finds a lower value of $\rho_0 = 270 (\pm 40)\,\mu\Omega$\,cm and a residual resistivity ratio RRR = 1.65, compared to  RRR $\approx$ 1.1 for ZrAs$_{1.58}$Se$_{0.39}$. This points to a similar, but somewhat less pronounced structural disorder in the free-of-vacancies phase. 
The magnetoresistance, which is a dimensionless relaxation-time-dependent quantity, provides further evidence for a far-reaching similarity between both Zr$Pn_xCh_y$ homologues. In fact, also for two ZrAs$_{1.58}$Se$_{0.39}$ specimens, neither any temperature dependence nor any deviation from the $B^{2}$ behavior was observed up to 20\,K, as shown in the inset of Fig.~3(b). From the values of MR $\propto (1/\rho)^2$ one can anticipate that  $\rho_0$ of sample \#2 excesses that of sample \#1 by about 18\,\%. This underlines that our estimation of a minor (12\,\%) difference between the measured $\rho_0$ values is accurate.

\section{III Screening factor, electron diffusion constants and other physical parameters of Z\lowercase{r}-based pnictide chalcogenides}

In order to estimate the screening factor $F$ for single crystalline ZrAs$_{1.58}$Se$_{0.39}$  and ZrP$_{1.54}$S$_{0.46}$, we turn to the Hartree-Fock approximation, following the 
formalism presented in \cite{Akkermans}. 
A measure of the strength of the electronic correlations is provided by the dimensionless parameter $a$, which is the ratio between Coulomb potential energy and kinetic energy. The parameter $a$ is thus proportional to the average distance between electrons $r_{\rm{s}}$, measured in multiples of the Bohr radius $a_0$. We estimate $a$ from
\begin{equation}
a=\left(\frac{r_{\rm{s}}}{a_0}\right)=\left(\frac{3}{4\pi n}\right)^{\frac{1}{3}}\frac{1}{a_0}
\end{equation}
where $n$ is the density of the electron gas in three dimensions. Taking the experimental value of $n\approx 0.45\times 10^{22}$\,cm$^{-3}$,  for the ferromagnetic isostructural system UAsSe~\cite{Reim.86},  we obtain $a = 7.1$. [It is worth to mentioning that UAsSe (nominal composition) with the Curie temperature of about 110\,K displays sample-dependent anomalies in the resistivity far below the ferromagnetic transition \cite{Cichorek2002}.]
Consequently, the ratio between the screening vector and the Fermi wavevector $k/k_{\rm{F}}$ is given by:
\begin{equation}
\frac{k}{k_{\rm{F}}}= \left(\frac{16}{3\pi^2}\right)^{\frac{1}{3}}\left(\frac{m*}{m_0}\right)^{\frac{1}{2}}\sqrt{a}.
\end{equation}

{
\normalsize{
\begin{table*}[t!]
\caption{Normal state and superconducting  properties of the ZrAs$_{1.58}$Se$_{0.39}$  and ZrP$_{1.54}$S$_{0.46}$  single crystals studied in this paper. Note that the transport property estimates are for transport along the $c$ direction.}
\label{tab:Tab1}
\begin{tabular}{p{0.4\textwidth}p{0.2\textwidth}p{0.2\textwidth}p{0.2\textwidth}}
\\[-2mm]
\hline \hline
\\[-2mm]
&	ZrAs$_{1.58}$Se$_{0.39}$ \#1 &	ZrAs$_{1.58}$Se$_{0.39}$  \#2 &	ZrP$_{1.54}$S$_{0.46}$ \\[1mm] 
\hline 
\\[-2mm]
Residual resistivity $\rho_0$ ($\mu\Omega$\,cm) &	490 ($\pm$40) &	550($\pm$40) &	270($\pm$40) \\
Residual resistivity ratio RRR & 	1.19 &	1.09 &	1.65 \\
Sommerfeld coefficient $\gamma$ (mJK$^{-2}$mol$^{-1}$) &	1.7 &	1.7 &	1.7 \\
Debye temperature $\Theta_{\rm{D}}$ (K) &	356 &	356 &	477 \\
Density of states $N(E_{\rm{F}})$ ($\times 10^{40}$ J$^{-1}$cm$^{-3}$) &	7.9 &	7.9 &	9.0 \\
Superconducting critical temperature $T_{\rm{c}}$ (K) &	0.13 &	0.14 &	3.7 \\
Upper critical field $B_{\rm{c}2}$ (T) &	$<$0.05 &	$<$0.05 &	$<$0.5 \\
Electron diffusion constant $D$ (cm$^2$s$^{-1}$) &	1.0 &	0.9  &	1.6 \\
Disorder parameter $k_{\rm{F}}l$  &	4.5 &	4.0 &	7.1 \\[1mm]
\hline\hline
\end{tabular}
\end{table*}}
}

The value of the effective mass $m^*$ in ZrAs$_{1.58}$Se$_{0.39}$  and ZrP$_{1.54}$S$_{0.46}$ can be estimated from 
the electronic specific-heat coefficient $\gamma$, which results in $k/k_{\rm{F}}= 2.06$ for both Zr-based pnictide chalcogenides.

For a 3D electron gas the screening factor $F$ is defined by  
\begin{equation}
F=\frac{k^2}{4k_{\rm{F}}^2}\ln\left(1+\frac{4k_{\rm{F}}^2}{k^2}\right).
\end{equation}

Within Thomas-Fermi theory one finds $F$ varying  between 1 for strong and $\approx 0$ for weak screening. For ZrAs$_{1.58}$Se$_{0.39}$  and ZrP$_{1.54}$S$_{0.46}$  one obtains $ F = 0.81$. Thus, $m^*$ and $n$ are compatible with  very efficient screening in ZrAs$_{1.58}$Se$_{0.39}$  and ZrP$_{1.54}$S$_{0.46}$. 

Finally, it has been pointed out~\cite{Altshuler1985,Lee1985} that in all formulas $F$ should be replaced by $\tilde F_{\sigma}$ given by 
\begin{equation}
\tilde{F}_\sigma(F)=-\frac{32 \left(-(0.5 F+1)^{3/2}+\frac{3 F}{4}+1\right)}{3 F},
\label{equ:EEI:Fs}
\end{equation}
if resistivity corrections due to EEI are calculated.
Note that for  very weak electron screening, i.e. $F\ll 1$, the expression (\ref{equ:EEI:Fs}) reduces to $\tilde F_{\sigma}\approx F$. 
For $F=0.81$, the value we obtained for  ZrAs$_{1.58}$Se$_{0.39}$  and ZrP$_{1.54}$S$_{0.46}$, $\tilde{F}_\sigma(F=0.81)\approx 0.76$.

In disordered metals with random scattering potential the electron motion is diffusive due to repeated scatterings with the  random potential.
The diffusive character of electron conduction is characterized by the diffusion constant $D$, which  can be estimated through the Einstein relation,
\begin{equation}
1/\rho_0={N(E_{\rm{F}})e^2}D,
\end{equation}
where $N(E_{\rm{F}})$ is the conduction electron density of states at the Fermi level and  $\rho_0$ is the residual resistivity. 
$N(E_{\rm{F}})$ can be estimated from the free-electron formula:
\begin{equation}
  N(E_{\rm{F}})=\frac{3\gamma}{\left(k_{\rm{B}}\pi\right)^2},
\label{eq:NEF}
\end{equation} 
where $\gamma$ is the Sommerfeld coefficient of the molar electronic specific heat  and $k_{\rm{B}}$ is the Boltzmann constant.  
For both pnictide chalcogenides, we have obtained similar values of  $N\left(E_{\rm{F}}\right)$, i.e., {approximately} $7.9\times 10^{40}$\,J$^{-1}$cm$^{-3}$ and $9.0\times 10^{40}$\,J$^{-1}$cm$^{-3}$ for ZrAs$_{1.58}$Se$_{0.39}$  and ZrP$_{1.54}$S$_{0.46}$, respectively. 
Consequently, the electron diffusion constant amounts to $D\approx 1.0$ and 1.6 cm$^2$s$^{-1}$ for the phase with and without vacant lattice sites, respectively.
The disorder parameter  $k_{\rm{F}}l$  (with $l$ being the mean free path and $ k_{\rm{F}}$  the Fermi wavenumber) turns out to be
$k_{\rm{F}}l \approx 4.2$ for ZrAs$_{1.58}$Se$_{0.39}$  and $k_{\rm{F}}l \approx 7.1$ for ZrP$_{1.54}$S$_{0.46}$.
This estimate has been obtained using the free-electron relation
 \begin{equation}
k_{\rm{F}}l=\frac{3m^{\star}D}{\hbar},
\end{equation}
where $m^{\star}\approx 1.7 m$ is the effective electron mass  obtained from the specific heat ($m$ being the bare electron mass).
Because of the small diffusion constants, a strong rather than weak disorder effect is anticipated for both Zr-based pnictide chalcogenides.
Table \ref{tab:Tab1} lists several  physical properties for both materials. 

Finally, we emphasize that our estimate of the residual resistivity  (cf.~Table \ref{tab:Tab1}) is accurate for the rather small samples of ZrAs$_{1.58}$Se$_{0.39}$  and ZrP$_{1.54}$S$_{0.46}$. This is inferred from the isothermal response of the resistivity to a magnetic field that was studied by taking advantage of the fact that the magnetoresistivity MR$\,=\,[(\rho(B)\,-\,\rho(0)]/\rho(0)$ is a dimensionless quantity. The MR data are depicted in the inset of Fig.~3(c) of the main text. Neither any temperature dependence nor any deviation from the $B^2$ behavior was observed in the temperature window 0.1\,K -- 20\,K and for $B > 1$\,T. In this field range, the Lorentz force leads to the curvature of the electronic trajectories and hence, a change of the classical transverse MR\,$\propto\,(B/\rho)^2$. At $B = 14$\,T, we have observed an MR of, respectively, 0.27\,\% and 0.18\,\%, i.e., a difference of about 35\,\% between the two specimens of ZrAs$_{1.58}$Se$_{0.39}$. 
Thus, one can anticipate that $\rho_0$ of sample \#2 exceeds that of sample \#1 by about 18\,\%. Note that this estimate, being in satisfactory agreement with the 12\,\% difference between the measured $\rho_0$ values, does not alter the discussion presented in Sec.~IV. For ZrP$_{1.54}$S$_{0.46}$  we have found that the MR along the $c$ axis is $1.4$\,\% at $B = 14$\,T. This is in accord with the lower residual resistivity compared to that  in ZrAs$_{1.58}$Se$_{0.39}$, which lends additional support to the similarity in basic physical properties of both Zr-based pnictide chalcogenides.  
%

\section{IV Analysis of the dephasing rates, negligible electron-electron interaction, and dynamic scattering centers}

This section contains our in-depth analysis  of the dephasing rates of ZrAs$_{1.58}$Se$_{0.39}$ and ZrP$_{1.54}$S$_{0.46}$.
It will lead us to the conclusion that dynamic scattering centers in the vacancy-containing pnictogen layer
of ZrAs$_{1.58}$Se$_{0.39}$ are responsible for the observed low-temperature transport anomalies in ZrAs$_{1.58}$Se$_{0.39}$.
Finally, we show that the dynamic scattering centers act as superconducting pair breakers. Thus, the difference in $T_c$ of the two systems is a consequence of this particular type of 2CK "impurities".

\subsection{Quantum interference effects in {ZrP$_{1.54}$S$_{0.46}$}  and dephasing in ZrAs$_{1.58}$Se$_{0.39}$}

In the absence of a magnetic field, superconducting fluctuations dominate the transport properties of ZrP$_{1.54}$S$_{0.46}$  leading to a strong decrease of $\rho(T)$ in a temperature region which significantly exceeds $T_{\rm{c}}\approx 3.7$\,K.  Under these conditions it is extremely difficult to convincingly separate corrections to the resistivity due to weak localization and EEI \cite{Altshuler1985,Lee1985}. 
Thus, we have performed experiments at finite magnetic field  to  suppress superconductivity and to explore the field dependence of the normal-state resistivity. In Fig.~\ref{fig:FigS2}, the relative change of the resistivity normalized to the minimum value, $(\rho-\rho_{\rm{min}})/\rho_{\rm{min}}$, is shown as a function of $T^{1/2}$ for fields up to 14\,T  (with field direction parallel to the current along the $c$ axis).  

At $B = 1$\,T a small fraction of the ZrP$_{1.54}$S$_{0.46}$  sample still displays surface superconductivity below $\approx$1\,K,
while at elevated temperatures, we observe an upturn in $\rho(T)$ which is more complex than the $-|A|T^{1/2}$ behavior 
found in ZrAs$_{1.58}$Se$_{0.39}$ and is ascribed to weak localization. 
Note that in three dimensions, the corrections to the resistivity coming from EEI and the weak-localization correction have different temperature dependencies.  In the case of the weak-localization correction, 
the temperature enters only through the relaxation rates for inelastic scattering processes. The standard way to experimentally extract the correction due to EEI is to apply a magnetic field in order to suppress the weak localization~\cite{Altshuler1985,Lee1985}.  The effect of an external magnetic field is to destroy the phase coherence of the partial electron waves. Since weak localization leads to an increase of the resistivity upon cooling, its suppression by a magnetic field  results in a negative magnetoresistivity. 

A negative MR in the ZrP$_{1.54}$S$_{0.46}$ compound is inferred from the $B\geq 1$\,T data (cf. Fig.~\ref{fig:FigS2}). Already at intermediate fields of about 2\,T, the $(\rho-\rho_{\rm{min}})/\rho_{\rm{min}}$ correction is smaller than for $B = 1$\,T. 
For the largest available field of 14\,T, the low-$T$ upturn is strongly reduced, and its magnitude is about 7 times smaller than for $B = 1$\,T. 
Several additional observations can be made: The application of a magnetic field shifts the position of $\rho_{\rm{min}}$ towards lower temperatures. As a result, at $B = 14$\,T the resistivity of ZrP$_{1.54}$S$_{0.46}$   is essentially temperature-independent down to liquid-helium temperature, and its rise is restricted to a rather narrow temperature window, i.e., from above 4.2\,K to around 1\,K. At $T< 1$\, K, the resistivity is found to saturate. (Similar complex behavior in the same temperature regime is already observed for $B = 2$\,T). Taken together, these experimental findings imply that the absolute value of the weak-localization MR increases upon cooling and amounts to about 0.02\,\%  at $T = 2$\,K.  Thus, it is  nearly two orders of magnitude  smaller than the classical MR (cf. the inset of Fig.~3(c) of the main text). Therefore, the total MR of  ZrP$_{1.54}$S$_{0.46}$    is essentially temperature independent at $T\leq 10$\,K and approaches 1.4\,\% at $B =14$\,T. 

The weak-localization correction to  $\rho(T)$ in ZrP$_{1.54}$S$_{0.46}$  is in striking contrast to the field-independent $-|A|T^{1/2}$ term in $\rho(T)$ of ZrAs$_{1.58}$Se$_{0.39}$. 
As we shall conclude below, this term must be caused by the presence of dynamic "impurities" that lead to two-channel Kondo physics. Since two-channel Kondo "impurities" act as dephasing centers for conduction electrons due to their dynamic nature, no weak-localization correction in $\rho(T)$ is expected.
Note that the strong dephasing in ZrAs$_{1.58}$Se$_{0.39}$  cannot be caused simply by the vacancies.
\begin{figure}[!h]
\centering
\includegraphics[width=0.5 \linewidth,keepaspectratio=true]{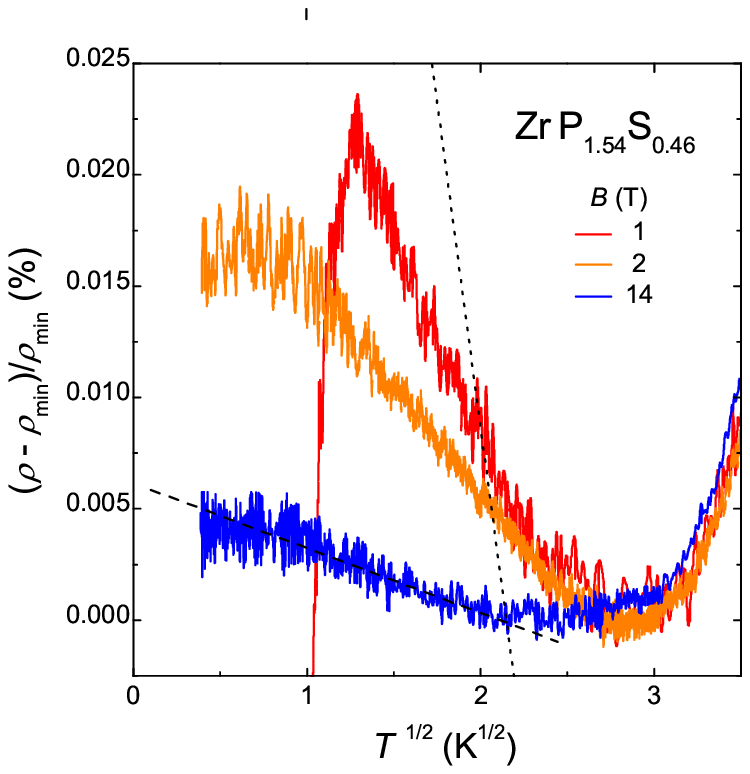}
\caption{
{\bf Low-temperature electrical resistivity of ZrP$_{1.54}$S$_{0.46}$}, as $(\rho-\rho_{\rm{min}})/\rho_{\rm{min}}$ vs. $T^{1/2}$, measured along $c$ axis in varying magnetic fields up to 14\,T. The slope of the dashed line is a prediction of the 3D EEI theory with $\tilde F_{\sigma}\approx 0$ and an electron diffusion constant as large as 647\,cm$^2$s$^{-1}$. The dotted line shows the magnitude of a hypothetical  $\tilde F_{\sigma}\approx 0$ EEI correction to the resistivity for a disordered metal with $\rho_0 = 270\,\mu\Omega$\,cm and $D = 1.6$\,cm$^2$s$^{-1}$, as experimentally found for ZrP$_{1.54}$S$_{0.46}$.
}
\label{fig:FigS2}
\end{figure}
Because of their static nature, vacancies act as elastic scatterer and do not destroy  phase coherence. However, this situation changes dramatically when the vacant sites trigger the formation of dynamic scattering centers. Quite generally, internal quantum degrees of freedom of  scatterers lead to dephasing in metals~\cite{Lin2002}. A well known example are magnetic impurities giving rise to spin-flip scattering~\cite{Micklitz2006}. If, for the specific case  of a Kondo impurity, the temperature is reduced to below the Kondo temperature, the dephasing time for the standard (single-channel) Kondo effect diverges for $T\to 0$. In contrast to (standard) magnetic Kondo impurities where the ground-state singlet has zero entropy, a doubly degenerate two-channel Kondo impurity in the non-Fermi liquid regime has a nonzero ground state entropy of $\frac{1}{2}$ln2. This implies that the dephasing time for the two-channel Kondo effect is finite even at $T = 0$ \cite{Zawadowski1999}.

The resistivity results for ZrP$_{1.54}$S$_{0.46}$ displayed in Fig.~\ref{fig:FigS2} lead to the conclusion that EEI very likely  does not contribute to the field-independent $-|A|T^{1/2}$ term in $\rho(T)$  of ZrAs$_{1.58}$Se$_{0.39}$:
For   ZrP$_{1.54}$S$_{0.46}$, the $\rho(T)$ upturn at $B$=14\,T is about 7 times smaller than the one at $B = 1$\,T. Within the EEI theory for three-dimensional disordered metals [see Eqs.~(\ref{eq:DrrEEI1})-(\ref{eq:DrrEEI4}) and discussion below], a field-independent correction to the resistivity requires  very weak screening ($\tilde F_{\sigma}\approx 0$) and consequently, a very large value of the diffusion constant, i.e., $D = 647$\, cm$^2$s$^{-1}$ (cf. dashed line in Fig.~\ref{fig:FigS2}). Estimates of the required  diffusion constants for such a field-independent contribution, based on Eq.~(\ref{eq:DrrEEI4}), are more than three orders of magnitude larger than the afore-mentioned values obtained from the Einstein relation. Furthermore, for a system with $\rho_0= 270\,\mu\Omega$\,cm and $D = 1.6$\,cm$^2$s$^{-1}$, like e.g. for  ZrP$_{1.54}$S$_{0.46}$, one would expect the $B$-independent $(\rho-\rho_{\rm{min}})/\rho_{\rm{min}}$ term to be as large as about 0.12 \%  at the lowest temperatures. Thus, if dominant, EEI would cause an easy-to-detect correction to the low-temperature resistivity, as indicated by the dotted line in Fig.~\ref{fig:FigS2}. Thus, we  conclude that the observed $-|A|T^{1/2}$ term in $\rho(T)$
of ZrAs$_{1.58}$Se$_{0.39}$  cannot be  caused by  EEI.
For more quantitative arguments, see the following subsection.

\subsection{Quantitative arguments for negligible electron-electron interactions  in ZrAs$_{1.58}$Se$_{0.39}$}
\label{SEC:EEI}

The only  mechanism, other than the proximity to a two-channel Kondo fixed point, that can result in a  $B$-independent $-|A|T^{1/2}$ correction to the resistivity  is the three-dimensional (3D) EEI in the so-called diffusion channel of disordered metals. This may occur in systems with an exceptionally  weak electron screening, i.e. $\tilde F_{\sigma}\approx 0$. 
In the presence of a magnetic field, the diffusion-channel correction to the resistivity can be written as a sum of two terms~\cite{Lee1985,Altshuler1985}:

\begin{equation}
\label{eq:DrrEEI1}
\left(\frac{\Delta\rho}{\rho}\right)_{\rm{EEI}}(B,T)=\left(\frac{\Delta\rho}{\rho }\right)_{\rm{EEI}}^{'}(T)+\left(\frac{\Delta\rho}{\rho}\right)_{\rm{EEI}}^{''}(B,T),
\end{equation}
where we assume that, in a metal at low temperatures, the corrections to the resistivity, $\delta \rho=\rho-\rho_0$, are small ($\delta \rho\ll \rho_0$).
The first term represents the field-independent exchange, $S_z = 0$ Hartree contribution, given by: 
\begin{equation}
\left(\frac{\Delta\rho}{\rho}\right)_{\rm{EEI}}^{'}\left(T\right)= -\rho\frac{0.919e^2}{4\pi^2\hbar}\left(\frac{4}{3}-\frac{1}{2}\tilde{F}_{\sigma}\right)\sqrt{\frac{k_{\rm{B}}T}{\hbar D}},
\label{eq:DrrEEI2}
\end{equation}
where $D$ is the electron diffusion constant and the screening factor $\tilde F_{\sigma}$ is defined in  Eq.~(\ref{equ:EEI:Fs}).
The second term in Eq.~(\ref{eq:DrrEEI1}) is the remaining $|S_z| = 1$ triplet contribution. Its field dependence is a result of Zeeman spin splitting and can be written as: 
\begin{equation}
\left(\frac{\Delta\rho}{\rho}\right)_{\rm{EEI}}^{''}\left(B,T\right)= \rho\frac{e^2}{4 \sqrt{2}\pi^2\hbar}g_3\left(h\right)\tilde{F}_{\sigma}\sqrt{\frac{k_{\rm{B}}T}{\hbar D}},
\label{eq:DrrEEI3}
\end{equation}
with $h =\frac{g\mu_{\rm{B}}B}{k_{\rm{B}}T}$, where $g$ is the Land\'{e} factor.  
Assuming $g = 2$, the possible values of  $g_3(h)$ at $B = 14$\,T are strongly $T$ dependent and hence vary between 12 to 0.2 when increasing temperature from 0.1\,to 10\,K (see Fig.~\ref{fig:FigS3}).

\begin{figure}[!h]
\begin{center}
\includegraphics[width=0.5 \linewidth,keepaspectratio=true]{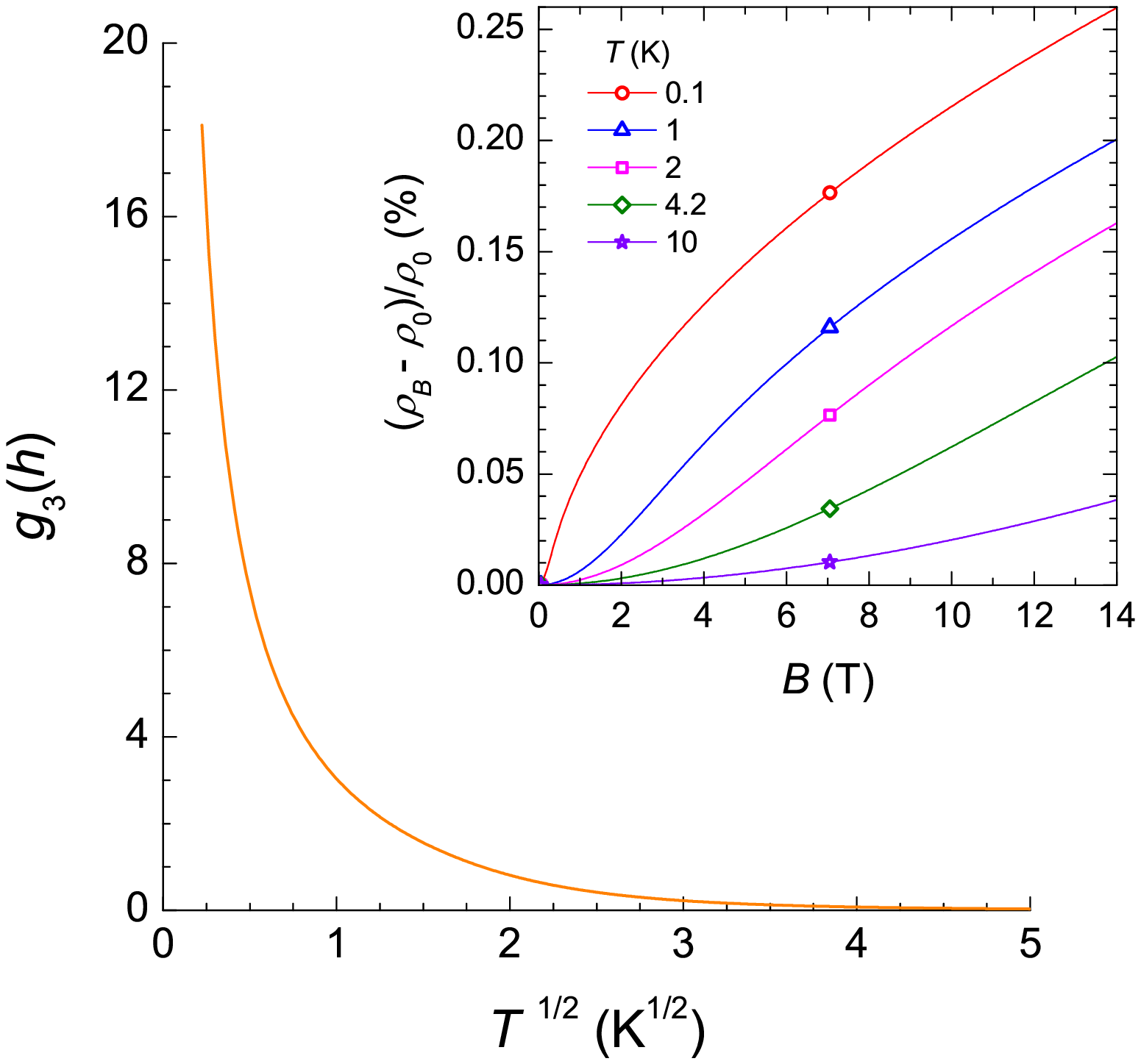}
\end{center}
\caption{
{\bf Temperature dependence of the $g_3(h)$ function.} Calculations, presented on a $T^{1/2}$ scale, were done for the constant field $B = 14$\,T. Inset: The diffusion-channel contribution to the magnetoresisitivity for various temperatures. $D = 0.91$\,cm$^2$s$^{-1}$, $\rho = 550\,\mu\Omega$\,cm,~$\tilde F_{\sigma}  = 0.72$, $g = 2$.
}
\label{fig:FigS3}
\end{figure}
In the absence of a magnetic field the EEI correction to the resistivity reduces to the following expression~\cite{Lee1985}:
\begin{equation}
\left(\frac{\Delta\rho}{\rho}\right)_{\textrm{EEI}}\left(0,T\right)= -\rho\frac{0.919e^2}{4\pi^2\hbar}\left(\frac{4}{3}-\frac{3}{2}\tilde{F}_{\sigma}\right)\sqrt{\frac{k_{\textrm{B}}T}{\hbar D}},
\label{eq:DrrEEI4}
\end{equation}
with contributions coming both from the singlet as well as the triplet term.
\begin{figure}[!ht]
\centering
\includegraphics[width=0.5 \linewidth,keepaspectratio=true]{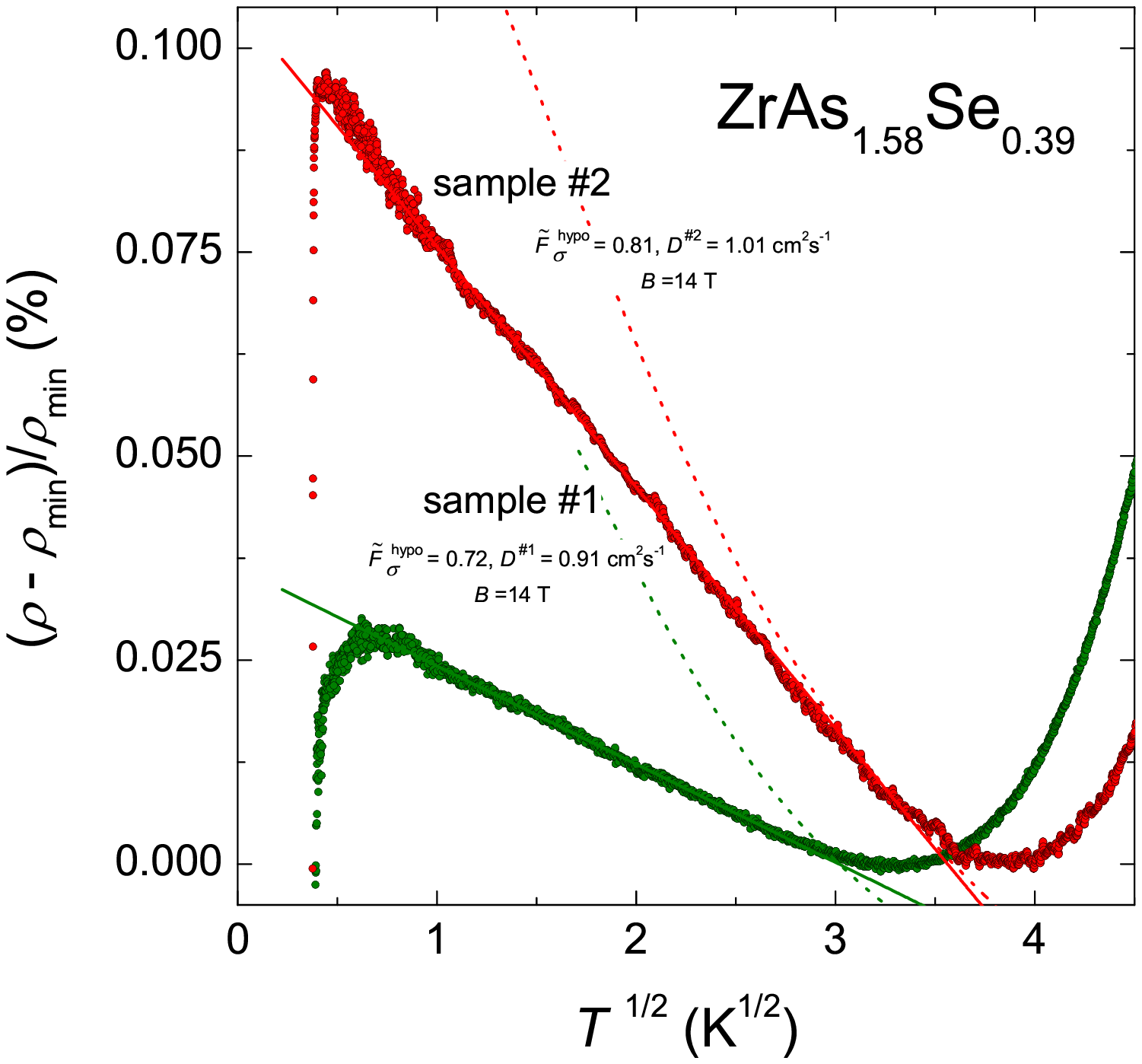}
\caption{{\bf Quantitative arguments for negligible EEI in the Zr-As-Se system.}
Temperature dependence of the normalized zero-field $\rho(T)$ for the ZrAs$_{1.58}$Se$_{0.39}$  samples as $(\rho-\rho_{\rm{min}})/\rho_{\rm{min}}$ vs. $T^{1/2}$. Continuous, straight lines display the 3D EEI theoretical correction for $\tilde F_{\sigma}^{\mbox{\tiny assumed}}=0$, which implies  the unrealistic values $D = 124$ cm$^2$s$^{-1}$ and 26 cm$^2$s$^{-1}$ for sample \#1 and \#2 at $B=0$, respectively.
The Einstein relation yields $D = 1.01$ cm$^2$s$^{-1}$ and 0.91 cm$^2$s$^{-1}$ instead. Dotted lines indicate the expected results at $B=14$\,T for $F_{\sigma}^{\mbox{\tiny hypo}}=0.71$ and $0.81$ respectively, if the $T^{1/2}$ term would be caused by EEI.
}
\label{fig:FigS4}
\end{figure}

It follows from Eq.~(\ref{eq:DrrEEI3}) that a $B$-independent $-|A|T^{1/2}$ correction to $\rho(T)$ is in principle possible if $\tilde{F}_\sigma\approx 0$.
According to Section III, theoretical values of  the corrected screening factor for single crystalline ZrAs$_{1.58}$Se$_{0.39}$  and ZrP$_{1.54}$S$_{0.46}$ are
 $\tilde F_{\sigma}\approx 0.66$. Therefore, it is inconsistent to attribute the  $B$-field independent  $-|A|T^{1/2}$ term in the resistivity of   ZrAs$_{1.58}$Se$_{0.39}$  to EEI. To further expand on this, we will now {\it assume} $\tilde F_{\sigma}^{\mbox{\tiny assumed}}\approx 0$ for  
ZrAs$_{1.58}$Se$_{0.39}$ and demonstrate that this is indeed inconsistent with our experimental results.

With this assumption, we have fitted the zero-field $\rho(T)$ data for samples \#1 ($\rho_{\#1} = 490\,\mu\Omega$\,cm) and \#2 ($\rho_{\#2} = 550\,\mu\Omega$\,cm) leaving $D$ as a free parameter. Satisfactory fits, which are displayed  as continuous, straight lines in Fig.~\ref{fig:FigS4},  require electron diffusion constants $D_{\#1} = 124$\,cm$^2$s$^{-1}$  and  $D_{\#2} = 26$\,cm$^2$s$^{-1}$, respectively.  However, such values are unrealistic for several reasons: First of all, a difference in electron diffusion constants by a factor of 6 is not expected between specimens with  very similar residual resistivities and densities of states at the Fermi level.  
Note that the same Sommerfeld coefficient $\gamma = 1.7$\,mJ/K$^2$mol as in ZrAs$_{1.58}$Se$_{0.39}$  was reported for ZrAs$_{1.40}$Se$_{0.50}$~\cite{Schmidt2005a}.
This implies a negligible variation of $N(E_{\rm{F}})$  inside the narrow homogeneity range of the tetragonal \mbox{Zr--As--Se} phase.  
Furthermore, an essentially sample-independent $N(E_{\rm{F}})$  is expected from the very similar superconducting transition temperatures $T_{\rm{c}}~\propto$~exp$[-1/N(E_{\rm{F}})]$ of the ZrAs$_{1.58}$Se$_{0.39}$  samples. 
Finally, estimates of diffusion constants based on Eq.~(\ref{eq:DrrEEI4}) are by up to 3 orders of magnitude larger than the corresponding values obtained from the Einstein relation: 
as shown in Section III, combining the resistivity and specific-heat results yields $D_{\#1} = 1.01$ cm$^2$s$^{-1}$ and  $D_{\#2} =  0.91$ cm$^2$s$^{-1}$. These values are comparable to typical values of the electron diffusion constant for $sp$-band amorphous metals  or  $d$-band amorphous alloys,  which amount to a few of tenths of cm$^2$s$^{-1}$~\cite{Lin2002}. 

Hence, we reach the important conclusion that the assumption of $\tilde F_{\sigma}\approx 0$ for ZrAs$_{1.58}$Se$_{0.39}$ leads to a non-physical variation of the diffusive motion in (single crystalline) ZrAs$_{1.58}$Se$_{0.39}$. In turn, it is also possible to fit the zero-field $\rho(T)$ data for samples \#1 ($\rho_{\#1} = 490\,\mu\Omega$\,cm and $D_{\#1} = 1.01$\,cm$^2$s$^{-1}$) and \#2 ($\rho_{\#2} = 550\,\mu\Omega$\,cm and $D_{\#2} = 0.91$\,cm$^2$s$^{-1}$) to Eq. (\ref{eq:DrrEEI4}) leaving $\tilde F_{\sigma}$ as a free parameter. Least-squares fits yield  $\tilde F_{\sigma}^{\mbox{\tiny hypo}} = 0.81$ and 0.72 for samples \#1 and \#2, respectively. In such a case of very efficient screening,  the magnetic-field-dependent part of the  $S_z = 1$ Hartree term is expected to dephase electrons  giving rise to a positive magnetoresistivity, whose magnitude significantly increases with decreasing temperature, as shown in the inset of Fig.~\ref{fig:FigS3}. Consequently, this would lead to large changes of the $-|A|T^{1/2}$ term in $\rho(T)$ (cf. the dotted lines in Fig.~\ref{fig:FigS4}).

\section{V Two-channel Kondo physics from dynamic structural scattering centers}
\label{SEC:2CK}

The results on ZrAs$_{1.58}$Se$_{0.39}$ and ZrP$_{1.54}$S$_{0.46}$ reported here together with our analysis of possible electron-electron interaction and weak localization corrections imply the existence of a dephasing mechanism associated with vacancies in the pnictogen layers that is compatible only with 2CK scattering centers.
Here, we argue that such scattering centers are possible due to the square lattice symmetry of the vacancy carrying As layer and the tendency of such
square nets with the PbFCl structure-type compounds to develop a dynamic Jahn-Teller distortion~\cite{Tremel1987}.

The (single-channel) Kondo effect occurs when magnetic impurities interact antiferromagnetically with delocalized electronic states. The interplay of Kondo screening and the Jahn-Teller distortion has been discussed by Gogolin~\cite{Gogolin.96} and the effect of a dynamical Jahn-Teller distortion in cage compounds like skutterudites has been addressed by Hotta~\cite{Hotta2006}.
In contrast, 2CK physics arises when the local moment is coupled to two identical fermionic baths. In this case, overscreening of the impurity degree of freedom occurs and even the strong-coupling fixed point becomes unstable. The system flows to some intermediate
effective coupling and displays non-Fermi liquid behavior with a non-vanishing zero-temperature entropy, a logarithmically increasing spin susceptibility, and square-root-in-temperature behavior of the scattering rate. The 2CK model,
has turned out to be extremely difficult to realize, as e.g. any channel-symmetry breaking terms will drive the model away from the non-Fermi liquid fixed point. A very interesting proposal to realize two-channel Kondo physics has been made by Zawadowski and Vlad\'{a}r~\cite{Vladar1983a,Vladar1983b,Zawadowski1980} to explain transport anomalies often seen in metallic glasses.
The model by Vlad\'{a}r and Zawadowski is based on atoms in double-well potentials that may tunnel from one minimum to the other. These two minima of the atomic energy give raise to a pseudospin variable. Direct tunneling of the atom between the two minima
necessarily splits the  degeneracy between the two states and thus corresponds to an effective magnetic field. 
The conduction electron assisted tunneling of the atom assumes the role of a hybridization term. The conduction electron spin does not participate in assisted tunneling processes and  thus gives rise to two degenerate scattering channels protected by time-reversal symmetry. This opens up the possibility for the two-level system to flow to the 2CK fixed point. For energies well below the associated Kondo temperature, the system would therefore display a $\sqrt{T}$ behavior  insensitive
to an applied magnetic fields as the Kondo scattering processes are non-magnetic in character~\cite{Vladar1983a,Vladar1983b,Zawadowski1980,Cox1998}. However, as pointed out by Aleiner {\it et al.}, this proposal  also implies the existence of an effective magnetic field in excesses the associated Kondo scale so that a two-level system may never reach the regime where 2CK physics ensues~\cite{Aleiner2002}.

It thus would seem that the existence of non-magnetic 2CK scattering centers in ZrAs$_{1.58}$Se$_{0.39}$ is unlikely although  our analysis strongly supports their existence. A way out of this conundrum is to consider dynamical defects with higher symmetry~\cite{Moustakas.97}. If the dynamic defect is compatible with a triangular or $C_3$ symmetry, there exists a doublet whose degeneracy is ensured by symmetry. In this case, the low-energy properties are governed by a fixed point that is identical to the two-channel Kondo fixed point up to irrelevant operators.
A difficulty with this proposal is that the doublet is usually higher in energy than the singlet. Under certain conditions, the doublet can become the ground state. For the case of a defect with triangular symmetry, this has been investigated in Ref.~\cite{Moustakas.97}. More recently, the case of an SU(3)-symmetric defect in a metal was investigated by perturbative and NRG methods and it was found that the level spacing between doublet and singlet  always renormalizes down as a result of  the coupling to the electronic bath~\cite{Arnold2007}. As a result, the doublet is renormalized below the singlet in a wide parameter range~\cite{Arnold2007}.\\

Here, we carry over the analysis of Refs.~\cite{Moustakas.97,Arnold2007} to vacancies in square nets of the PbFCl structure-type compounds and show that the vacancies in the pnictogen layer give rise to dynamic defects with a $C_4$ symmetry. The square symmetry implies the existence of doubly generate eigenstates  that form an irreducible representation of the group $C_4$. The comparably small composition range $1.90\leq x+y \leq 1.99$ for which two-channel Kondo physics has been observed in ZrAs$_x$Se$_y$ implies that the vacancies in the pnictogen layer are in the dilute limit. 
Furthermore, as discussed above (see Sec.~I), the pnictogen layer $Pn$(2$a$)-site is only occupied up to 97\,\% with As.  Interstitial As does not occur in this layer. Thus, each vacancy preserves the $C_4$ symmetry of the pnictogen layer. As a result, a Jahn-Teller distortion forms in the pnictogen layer concomitant with a formation of As dimers or oligomers. This phenomenon is well known to occur in such square nets of the PbFCl structure-type compounds~\cite{Tremel1987}. The flattened displacement ellipsoids shown in Fig.~2(b)  are indicative of the occurrence of the  dynamic Jahn-Teller effect in the As (2$a$) layer. As the Jahn-Teller distortion develops, the doublet states split and one of them becomes the new ground state. A finite tunneling rate between the different impurity positions compatible with the overall square symmetry restores the square symmetry through a dynamic Jahn-Teller effect. The group $C_4$ possesses only one-dimensional irreducible representations and one two-dimensional irreducible representation (IRREP).
Thus, a non-Kramers doublet associated with the dynamic Jahn-Teller distortion transforms as the two-dimensional irreducible representation of the group $C_4$, and allows for a two-channel Kondo fixed point to occur at sufficiently low energies~\cite{Moustakas.97}. In this two-dimensional subspace, the basis states (labelled $\pm$) allow to introduce a  pseudospin label. Expanding the conduction electron states around the quantum impurity yields 
\begin{eqnarray}
H\,&=&\,H_{\mbox{\small loc}}\\ \nonumber
&+& Q_1\sum_{\sigma}(d^{\dagger}_{+}d^{}_{+}-d^{\dagger}_{-}d^{}_{-})(c^{\dagger}_{+,\sigma}c^{}_{+,\sigma}-c^{\dagger}_{-,\sigma}c^{}_{-,\sigma})\\ \nonumber
&+& \Delta_1 \sum_{\sigma} (d_{+}^{\dagger}d_{-}^{}c^{\dagger}_{-,\sigma}c^{}_{+,\sigma}+ d_{-}^{\dagger}d_{+}^{}c^{\dagger}_{+,\sigma}c^{}_{-,\sigma})\\ \nonumber
&+& H_{\mbox{\tiny additional}},
\end{eqnarray}
where $H_{\mbox{\small loc}}$ contains the local part of the dynamic defect, $d_{\pm}^{\dagger}$ creates an electron in the basis state $\pm$,   
$c^{\dagger}_{\pm,\sigma}$ is the local (energy integrated) conduction electron creation operator projected 
onto the basis states of the set of irreducible representations of the local symmetry using the great orthogonality theorem to obtain the invariant coupling between the local doublet subspace and the conduction electrons.  
The term proportional to $Q_1$ describes the coupling of the $z$-component of the pseudospins wheres $\Delta_1$ is the (pseudo-) spin-flip component responsible for Kondo-scattering processes and $H_{\mbox{\tiny additional}}$ contains all additional terms in the Hamiltonian (see Ref.~\cite{Moustakas.97} for a discussion of their relevance/irrelevance near the two-channel Kondo fixed point). As the conduction electron spin degree of freedom $\sigma$ only enters as an overall summation index and the degeneracy in $\sigma$ is protected by time-reversal symmetry, the model is equivalent to the two-channel Kondo model. A magnetic field breaks the spin degeneracy of the conduction electrons but this only leads to observable effects  for a field strength comparable to the bandwidth. That a magnetic field up to 14~Tesla does not affect the anomalous low-temperature behavior of  ZrAs$_{1.58}$Se$_{0.39}$  is a strong indication for two-channel physics. 

The Kondo temperature scale below which one observes 2CK physics in ZrAs$_{1.58}$Se$_{0.39}$ is comparatively large. This points to a correspondingly large tunneling rate associated with the dynamic defect. This is possible if the associated effective mass is small or the coupling between the As electrons and the conduction electrons is strong enough.  
The formation of As dimers or oligomers associated with the distortion does imply that the coupling between the As and conduction electrons can be large and
can result in a tunneling potential of small effective mass.

\section{VI Pair breaking due to dynamic structural scattering centers}
\label{SEC:pairbreaker}

Both homologues Zr$Pn_xCh_y$ develop superconductivity 
at sufficiently low temperatures, i.e. below $T_c\approx$~0.14\,K for ZrAs$_{1.58}$Se$_{0.39}$ and $T_{\rm{c}}\approx$~3.7\,K for ZrP$_{1.54}$S$_{0.46}$ (see Table  \ref{tab:Tab1}).
These compounds are standard (BCS) singlet superconductors with fully gapped Fermi surfaces, as can be inferred e.g.
from the exponential behavior of the heat capacity in the superconducting state [cf. Fig.~2(c) of the main text].
Despite their similar physical and chemical properties, the transition temperatures of both compounds differ  
by roughly a factor of $30$. This is unexpected. We show here, that this difference is a consequence of the presence of dynamic scattering centers in ZrAs$_{1.58}$Se$_{0.39}$.

For weak-coupling superconductors, $T_{\rm{c}}$ is related to  the
 superconducting gap at zero temperature, $\Delta(T=0)$
\begin{equation}
 T_c \sim \Delta(T=0),
\end{equation}
which, within BCS theory, obeys
\begin{equation}
 \Delta(T=0)=2\hbar \omega_{\rm{D}} \exp\Big [-\frac{1}{N(E_{\rm{F}}) V} \Big],
\end{equation}
where $\omega_{\rm{D}}$ is the Debye frequency, $N(E_{\rm{F}})$ is the density of states and $V$ is the strength of the net
attractive coupling. 
The Sommerfeld coefficients for  ZrAs$_{1.58}$Se$_{0.39}$ and ZrP$_{1.54}$S$_{0.46}$ are virtually identical
so that $N(E_{\rm{F}})$ of both compounds must be very similar. 
The difference  in the phonon contribution to the heat capacity
mainly reflects the difference in the atomic weights of the constituting elements of the two compounds, e.g. As and Se vs. P and S, and thus should result in only a moderate variation of $T_{\rm{c}}$.
The strong suppression of $T_{\rm{c}}$ in ZrAs$_{1.58}$Se$_{0.39}$ as compared to ZrP$_{1.54}$S$_{0.46}$ and its variation across samples from the same sample growth (see Table  \ref{tab:Tab1}) 
is another indication that dynamic, non-magnetic quantum impurities exist and that the low-energy behavior  in the ZrAs$_{1.58}$Se$_{0.39}$ compound is in line with 2CK physics, as we will now demonstrate. Note that according to Anderson's theorem, neither $\Delta$ nor $T_{\rm{c}}$ of $s$-wave superconductors are affected by the presence of weak disorder~\cite{Abrikosov1961}. (This theorem does not apply when magnetic impurities are present which act as pair-breakers.)

To address the effect of Cooper pair scattering off the dynamic impurity with $C_4$ symmetry, we model the superconducting
host  by the BCS Hamiltonian,
\begin{equation}
 H_0^{\mbox{\scriptsize BCS}}=\sum_{k,\sigma} \epsilon_k c^{\dagger}_{k,\sigma}c^{ }_{k,\sigma}-\Delta\sum_{k,\sigma} 
\Big(  c^{\dagger}_{k,\sigma}c^{\dagger}_{-k,-\sigma} + \mbox{h.c.} \Big),
\end{equation}
where (within BCS)
$\Delta=-V\sum_{k,\sigma}\langle c_{-k,-\sigma} c_{k,\sigma}\rangle$.
The order parameter describes the pairing of time-reversed conduction electron states. The 2CK effect in ZrAs$_{1.58}$Se$_{0.39}$ has been argued to arise out of the non-Kramers doublet transforming as the two-dimensional IRREP of $C_4$. This representation cannot be transformed into its complex conjugate through the application of a unitary transformation and thus the associated basis states of the two-dimensional IRREP, labeled by $+$ and $-$, are time-reversed partners. 
Expanding $\Delta$ around the quantum defect will thus have to involve singlets of $+$ and $-$. The full Hamiltonian including the quantum defect involves scattering from one of the basis states of the two-dimensional IRREP to the other. These scattering processes therefore have to break up Cooper pairs  
and thus reduces $T_{\rm{c}}$.

A more explicit demonstration that dynamic tunneling centers act as pair-breakers has e.g. been given in Ref.~\cite{Sellier2001}:
We model the dynamic  scattering center by an Anderson impurity model~\cite{Cox1998}
\begin{equation}
\label{Hand}
 H_{\mbox{\small 2CK}}=\sum_{m=\pm} \epsilon_f f_{m}^{\dagger}f^{}_{m}+\Gamma \sum_{\sigma,m}\Big [ c^{\dagger}_{\sigma,m}b^{\dagger}_{-\sigma}f^{}_{m}
+ \mbox{h.c.}\Big],
\end{equation}
where $m=\pm$ labels the two basis states of the two-dimensional IRREP.
In the representation underlying Eq.~(\ref{Hand}), a constraint
\begin{equation}
 Q=\sum_{m}f_{m}^{\dagger}f^{}_{m} +\sum_{\sigma} b^{\dagger}_{\sigma}b^{}_{\sigma} = 1,
\end{equation}
has to hold at all times.
$\Gamma$ is a matrix element characterizing the strength of the electron-assisted tunneling.
  Expanding the conduction electron operators around the dynamic scattering center in terms of spherical
harmonics 
\begin{equation}
 \sum_{k}c_{k,\sigma}^{\dagger}e^{i{\bf kr}}=\frac{1}{r}\sum_{l=0}^{\infty}\sum_{m=-l}^{l}
a_{l,m,\sigma}^{\dagger}({\bf r})Y_{lm}({\bf r}/r)
\end{equation}
and transforming back to (radial) momentum 
\begin{equation}
 c_{l,m,\sigma}(k)=(-i)^l\int dr kr j_l(kr) a_{l,m,\sigma}^{\dagger}({\bf r})
\end{equation}
where $j_l(kr)$ is the $l$th Bessel function of the first kind.
Inserting this expansion into the equation for the superconducting order parameter yields
\begin{equation}
 \Delta=-V\sum_{l,m,\sigma}\int_0^\infty dk (-1)^{l+m} \langle c_{l,m,\sigma}(k) c_{l,-m,-\sigma}(k)\rangle.
\end{equation}

As Cooper pairs are singlets in spin and pseudospin space, the quantum impurity will break Cooper pairs via electron
assisted tunneling, i.e. through flipping the pseudospin of an electron in much the same 
way as magnetic quantum impurities. As a consequence, Anderson's theorem does not apply.

For ZrAs$_{1.58}$Se$_{0.39}$, with a small number of such structural dynamic scattering centers, bulk superconductivity still sets in but at much lower
temperature than it would without such quantum impurities.  It is worth noting that here, the pseudospin is not screened thus giving rise
to non-Fermi liquid behavior and the dynamic defects remain pair breaking at lowest temperatures. This is in contrast to
Kondo impurities, which obey Anderson's theorem in the limit where the Kondo temperature is much larger than
the superconducting transition temperature.

\section{VII Temperature dependence of $\rho(T)$ above the $-|A| \sqrt{T}$ regime}
\label{SEC:Tbehav}

\begin{figure}[!ht]
\centering
\includegraphics[width=0.5 \linewidth,keepaspectratio=true]{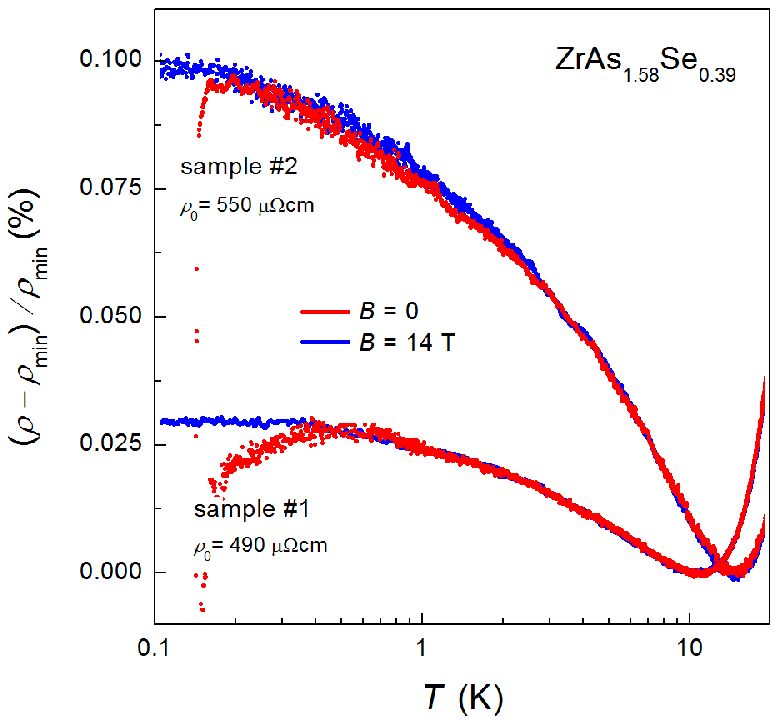}
\caption{ {\bf Low-temperature resistivity $(\rho(T) - \rho_{\mbox{\tiny min}})/\rho_{\mbox{\tiny min}}$ vs. $T$} }
\label{fig:highT}
\end{figure}

A logarithmic temperature dependence of the resistivity is expected for the 2CK at temperatures above the $T^{1/2}$ behavior. This will be overshadowed by the increase in resistivity due to additional scattering processes, i.e. scattering due to phonons. As a result, the log(T) behavior is observed only in a limited temperature range right below the temperature minimum.

\newpage
\end{widetext}


\begin{thebibliography}
\expandafter\ifx\csname natexlab\endcsname\relax\def\natexlab#1{#1}\fi
\expandafter\ifx\csname bibnamefont\endcsname\relax
  \def\bibnamefont#1{#1}\fi
\expandafter\ifx\csname bibfnamefont\endcsname\relax
  \def\bibfnamefont#1{#1}\fi
\expandafter\ifx\csname citenamefont\endcsname\relax
  \def\citenamefont#1{#1}\fi
\expandafter\ifx\csname url\endcsname\relax
  \def\url#1{\texttt{#1}}\fi
\expandafter\ifx\csname urlprefix\endcsname\relax\def\urlprefix{URL }\fi
\providecommand{\bibinfo}[2]{#2}
\providecommand{\eprint}[2][]{\url{#2}}

\bibitem[{\citenamefont{Kamihara et~al.}(2008)\citenamefont{Kamihara, Watanabe,
  Hirano, and Hosono}}]{Kamihara.08}
\bibinfo{author}{\bibfnamefont{Y.}~\bibnamefont{Kamihara}},
  \bibinfo{author}{\bibfnamefont{T.}~\bibnamefont{Watanabe}},
  \bibinfo{author}{\bibfnamefont{M.}~\bibnamefont{Hirano}}, \bibnamefont{and}
  \bibinfo{author}{\bibfnamefont{H.}~\bibnamefont{Hosono}},
  \bibinfo{journal}{J.~Am.~Chem.~Soc.} \textbf{\bibinfo{volume}{130}},
  \bibinfo{pages}{3296} (\bibinfo{year}{2008}).

\bibitem[{\citenamefont{Sato et~al.}(2009)\citenamefont{Sato, Sugawara, Aoki,
  and Harima}}]{Sato}
\bibinfo{author}{\bibfnamefont{H.}~\bibnamefont{Sato}},
  \bibinfo{author}{\bibfnamefont{H.}~\bibnamefont{Sugawara}},
  \bibinfo{author}{\bibfnamefont{Y.}~\bibnamefont{Aoki}}, \bibnamefont{and}
  \bibinfo{author}{\bibfnamefont{H.}~\bibnamefont{Harima}},
  \emph{\bibinfo{title}{Handbook of Magnetic Materials}}
  (\bibinfo{publisher}{Elsevier}, \bibinfo{address}{Amsterdam},
  \bibinfo{year}{2009}), chap. \bibinfo{chapter}{Magnetic properties of filled
  skutterudites}.

\bibitem[{\citenamefont{Shi et~al.}(2011)\citenamefont{Shi, Yang, Salvador,
  Chi, Cho, Wang, Bai, Yang, Zhang, and Chen}}]{Shi.11}
\bibinfo{author}{\bibfnamefont{X.}~\bibnamefont{Shi}},
  \bibinfo{author}{\bibfnamefont{J.}~\bibnamefont{Yang}},
  \bibinfo{author}{\bibfnamefont{J.~R.} \bibnamefont{Salvador}},
  \bibinfo{author}{\bibfnamefont{M.}~\bibnamefont{Chi}},
  \bibinfo{author}{\bibfnamefont{J.~Y.} \bibnamefont{Cho}},
  \bibinfo{author}{\bibfnamefont{H.}~\bibnamefont{Wang}},
  \bibinfo{author}{\bibfnamefont{S.}~\bibnamefont{Bai}},
  \bibinfo{author}{\bibfnamefont{J.}~\bibnamefont{Yang}},
  \bibinfo{author}{\bibfnamefont{W.}~\bibnamefont{Zhang}}, \bibnamefont{and}
  \bibinfo{author}{\bibfnamefont{L.}~\bibnamefont{Chen}},
  \bibinfo{journal}{J.~Am.~Chem.~Soc.} \textbf{\bibinfo{volume}{133}},
  \bibinfo{pages}{7837} (\bibinfo{year}{2011}).

\bibitem[{\citenamefont{Hsieh et~al.}(2008)\citenamefont{Hsieh, Qian, Wray,
  Xia, Hor, Cava, and Hasan}}]{Hsieh.08}
\bibinfo{author}{\bibfnamefont{D.}~\bibnamefont{Hsieh}},
  \bibinfo{author}{\bibfnamefont{D.}~\bibnamefont{Qian}},
  \bibinfo{author}{\bibfnamefont{L.}~\bibnamefont{Wray}},
  \bibinfo{author}{\bibfnamefont{Y.}~\bibnamefont{Xia}},
  \bibinfo{author}{\bibfnamefont{Y.~S.} \bibnamefont{Hor}},
  \bibinfo{author}{\bibfnamefont{R.~J.} \bibnamefont{Cava}}, \bibnamefont{and}
  \bibinfo{author}{\bibfnamefont{M.~Z.} \bibnamefont{Hasan}},
  \bibinfo{journal}{Nature} \textbf{\bibinfo{volume}{452}},
  \bibinfo{pages}{970} (\bibinfo{year}{2008}).

\bibitem[{\citenamefont{Liu et~al.}(2014)\citenamefont{Liu, Jiang, Zhou, Wang,
  Zhang, Weng, Prabhakaran, Mo, Peng, Dudin et~al.}}]{Liu.14}
\bibinfo{author}{\bibfnamefont{Z.~K.} \bibnamefont{Liu}},
  \bibinfo{author}{\bibfnamefont{J.}~\bibnamefont{Jiang}},
  \bibinfo{author}{\bibfnamefont{B.}~\bibnamefont{Zhou}},
  \bibinfo{author}{\bibfnamefont{Z.~J.} \bibnamefont{Wang}},
  \bibinfo{author}{\bibfnamefont{Y.}~\bibnamefont{Zhang}},
  \bibinfo{author}{\bibfnamefont{H.~M.} \bibnamefont{Weng}},
  \bibinfo{author}{\bibfnamefont{D.}~\bibnamefont{Prabhakaran}},
  \bibinfo{author}{\bibfnamefont{S.-K.} \bibnamefont{Mo}},
  \bibinfo{author}{\bibfnamefont{H.}~\bibnamefont{Peng}},
  \bibinfo{author}{\bibfnamefont{P.}~\bibnamefont{Dudin}},
  \bibnamefont{et~al.}, \bibinfo{journal}{Nat.~Mater.}
  \textbf{\bibinfo{volume}{13}}, \bibinfo{pages}{677} (\bibinfo{year}{2014}).

\bibitem[{\citenamefont{Xu et~al.}(2015)\citenamefont{Xu, Alidoust, Belopolski,
  Yuan, Bian, Chang, Zheng, Strocov, Sanchez, Chang et~al.}}]{Xu.15}
\bibinfo{author}{\bibfnamefont{S.-Y.} \bibnamefont{Xu}},
  \bibinfo{author}{\bibfnamefont{N.}~\bibnamefont{Alidoust}},
  \bibinfo{author}{\bibfnamefont{I.}~\bibnamefont{Belopolski}},
  \bibinfo{author}{\bibfnamefont{Z.}~\bibnamefont{Yuan}},
  \bibinfo{author}{\bibfnamefont{G.}~\bibnamefont{Bian}},
  \bibinfo{author}{\bibfnamefont{T.-R.} \bibnamefont{Chang}},
  \bibinfo{author}{\bibfnamefont{H.}~\bibnamefont{Zheng}},
  \bibinfo{author}{\bibfnamefont{V.~N.} \bibnamefont{Strocov}},
  \bibinfo{author}{\bibfnamefont{D.~S.} \bibnamefont{Sanchez}},
  \bibinfo{author}{\bibfnamefont{G.}~\bibnamefont{Chang}},
  \bibnamefont{et~al.}, \bibinfo{journal}{Nat.~Phys.}
  \textbf{\bibinfo{volume}{11}}, \bibinfo{pages}{748} (\bibinfo{year}{2015}).

\bibitem[{\citenamefont{Aleiner and Controzzi}(2002)}]{Aleiner2002}
\bibinfo{author}{\bibfnamefont{I.}~\bibnamefont{Aleiner}} \bibnamefont{and}
  \bibinfo{author}{\bibfnamefont{D.}~\bibnamefont{Controzzi}},
  \bibinfo{journal}{Physical Review B} \textbf{\bibinfo{volume}{66}},
  \bibinfo{pages}{045107} (\bibinfo{year}{2002}).

\bibitem[{\citenamefont{Moustakas and
  Fisher}(1996{\natexlab{a}})}]{Moustakas96}
\bibinfo{author}{\bibfnamefont{A.~L.} \bibnamefont{Moustakas}}
  \bibnamefont{and} \bibinfo{author}{\bibfnamefont{D.~S.}
  \bibnamefont{Fisher}}, \bibinfo{journal}{Phys.~Rev.~B}
  \textbf{\bibinfo{volume}{53}}, \bibinfo{pages}{4300}
  (\bibinfo{year}{1996}{\natexlab{a}}).

\bibitem[{\citenamefont{Hewson}(1993)}]{Hewson}
\bibinfo{author}{\bibfnamefont{A.~C.} \bibnamefont{Hewson}},
  \emph{\bibinfo{title}{The {K}ondo {P}roblem to {H}eavy {F}ermions}}
  (\bibinfo{publisher}{Cambridge University Press},
  \bibinfo{address}{Cambridge}, \bibinfo{year}{1993}).

\bibitem[{\citenamefont{Nozi\`{e}res and Blandin}(1980)}]{Nozieres.80}
\bibinfo{author}{\bibfnamefont{P.}~\bibnamefont{Nozi\`{e}res}}
  \bibnamefont{and} \bibinfo{author}{\bibfnamefont{A.}~\bibnamefont{Blandin}},
  \bibinfo{journal}{J.~Phys.} \textbf{\bibinfo{volume}{41}},
  \bibinfo{pages}{193} (\bibinfo{year}{1980}).

\bibitem[{\citenamefont{Zawadowski}(1980)}]{Zawadowski1980}
\bibinfo{author}{\bibfnamefont{A.}~\bibnamefont{Zawadowski}},
  \bibinfo{journal}{Physical Review Letters} \textbf{\bibinfo{volume}{45}},
  \bibinfo{pages}{211} (\bibinfo{year}{1980}).

\bibitem[{\citenamefont{Vlad\'{a}r and
  Zawadowski}(1983{\natexlab{a}})}]{Vladar1983a}
\bibinfo{author}{\bibfnamefont{K.}~\bibnamefont{Vlad\'{a}r}} \bibnamefont{and}
  \bibinfo{author}{\bibfnamefont{A.}~\bibnamefont{Zawadowski}},
  \bibinfo{journal}{Physical Review B} \textbf{\bibinfo{volume}{28}},
  \bibinfo{pages}{1596} (\bibinfo{year}{1983}{\natexlab{a}}).

\bibitem[{\citenamefont{Vlad\'{a}r and
  Zawadowski}(1983{\natexlab{b}})}]{Vladar1983b}
\bibinfo{author}{\bibfnamefont{K.}~\bibnamefont{Vlad\'{a}r}} \bibnamefont{and}
  \bibinfo{author}{\bibfnamefont{A.}~\bibnamefont{Zawadowski}},
  \bibinfo{journal}{Physical Review B} \textbf{\bibinfo{volume}{28}},
  \bibinfo{pages}{1564} (\bibinfo{year}{1983}{\natexlab{b}}).

\bibitem[{\citenamefont{Cochrane et~al.}(1975)\citenamefont{Cochrane, Harris,
  Str\"{o}m-Olson, and Zuckermann}}]{Cochrane1975}
\bibinfo{author}{\bibfnamefont{R.}~\bibnamefont{Cochrane}},
  \bibinfo{author}{\bibfnamefont{R.}~\bibnamefont{Harris}},
  \bibinfo{author}{\bibfnamefont{J.}~\bibnamefont{Str\"{o}m-Olson}},
  \bibnamefont{and}
  \bibinfo{author}{\bibfnamefont{M.}~\bibnamefont{Zuckermann}},
  \bibinfo{journal}{Physical Review Letters} \textbf{\bibinfo{volume}{35}},
  \bibinfo{pages}{676} (\bibinfo{year}{1975}).

\bibitem[{\citenamefont{Ralph et~al.}(1994)\citenamefont{Ralph, Ludwig, von
  Delft, and Buhrman}}]{Ralph1994}
\bibinfo{author}{\bibfnamefont{D.}~\bibnamefont{Ralph}},
  \bibinfo{author}{\bibfnamefont{A.}~\bibnamefont{Ludwig}},
  \bibinfo{author}{\bibfnamefont{J.}~\bibnamefont{von Delft}},
  \bibnamefont{and} \bibinfo{author}{\bibfnamefont{R.}~\bibnamefont{Buhrman}},
  \bibinfo{journal}{Physical Review Letters} \textbf{\bibinfo{volume}{72}},
  \bibinfo{pages}{1064} (\bibinfo{year}{1994}).

\bibitem[{\citenamefont{Halbritter et~al.}(2000)\citenamefont{Halbritter,
  Kolesnychenko, {Mih\'{a}ly}, Shklyarevskii, and van Kempen}}]{Halbritter.00}
\bibinfo{author}{\bibfnamefont{A.}~\bibnamefont{Halbritter}},
  \bibinfo{author}{\bibfnamefont{O.~Y.} \bibnamefont{Kolesnychenko}},
  \bibinfo{author}{\bibfnamefont{G.}~\bibnamefont{{Mih\'{a}ly}}},
  \bibinfo{author}{\bibfnamefont{O.~I.} \bibnamefont{Shklyarevskii}},
  \bibnamefont{and} \bibinfo{author}{\bibfnamefont{H.}~\bibnamefont{van
  Kempen}}, \bibinfo{journal}{Phys.~Rev.~B} \textbf{\bibinfo{volume}{61}},
  \bibinfo{pages}{1564} (\bibinfo{year}{2000}).

\bibitem[{\citenamefont{Huang et~al.}(2007)\citenamefont{Huang, Lee, Akimoto,
  Kono, and Lin}}]{Huang2007}
\bibinfo{author}{\bibfnamefont{S.}~\bibnamefont{Huang}},
  \bibinfo{author}{\bibfnamefont{T.}~\bibnamefont{Lee}},
  \bibinfo{author}{\bibfnamefont{H.}~\bibnamefont{Akimoto}},
  \bibinfo{author}{\bibfnamefont{K.}~\bibnamefont{Kono}}, \bibnamefont{and}
  \bibinfo{author}{\bibfnamefont{J.}~\bibnamefont{Lin}},
  \bibinfo{journal}{Physical Review Letters} \textbf{\bibinfo{volume}{99}},
  \bibinfo{pages}{046601} (\bibinfo{year}{2007}).

\bibitem[{\citenamefont{Zhu et~al.}(2016)\citenamefont{Zhu, Nie, Xiong,
  Schlottmann, and Zhao}}]{Zhu.16}
\bibinfo{author}{\bibfnamefont{L.~J.} \bibnamefont{Zhu}},
  \bibinfo{author}{\bibfnamefont{S.~H.} \bibnamefont{Nie}},
  \bibinfo{author}{\bibfnamefont{P.}~\bibnamefont{Xiong}},
  \bibinfo{author}{\bibfnamefont{P.}~\bibnamefont{Schlottmann}},
  \bibnamefont{and} \bibinfo{author}{\bibfnamefont{J.~H.} \bibnamefont{Zhao}},
  \bibinfo{journal}{Nat.~Commun.} \textbf{\bibinfo{volume}{7}},
  \bibinfo{pages}{10817} (\bibinfo{year}{2016}).


\bibitem[{\citenamefont{Aleiner et~al.}(2001)\citenamefont{Aleiner, Altshuler,
  Galperin, and Shutenko}}]{Aleiner2001}
\bibinfo{author}{\bibfnamefont{I.}~\bibnamefont{Aleiner}},
  \bibinfo{author}{\bibfnamefont{B.}~\bibnamefont{Altshuler}},
  \bibinfo{author}{\bibfnamefont{Y.}~\bibnamefont{Galperin}}, \bibnamefont{and}
  \bibinfo{author}{\bibfnamefont{T.}~\bibnamefont{Shutenko}},
  \bibinfo{journal}{Physical Review Letters} \textbf{\bibinfo{volume}{86}},
  \bibinfo{pages}{2629} (\bibinfo{year}{2001}).

\bibitem[{\citenamefont{Zar\'{a}nd}(2005)}]{Zarand2005}
\bibinfo{author}{\bibfnamefont{G.}~\bibnamefont{Zar\'{a}nd}},
  \bibinfo{journal}{Physical Review B} \textbf{\bibinfo{volume}{72}},
  \bibinfo{pages}{245103} (\bibinfo{year}{2005}).

\bibitem[{\citenamefont{Altshuler and Aronov}(1985)}]{Altshuler1985}
\bibinfo{author}{\bibfnamefont{B.}~\bibnamefont{Altshuler}} \bibnamefont{and}
  \bibinfo{author}{\bibfnamefont{A.}~\bibnamefont{Aronov}},
  \emph{\bibinfo{title}{{Electron-Electron Interactions in Disordered
  Systems}}} (\bibinfo{publisher}{North-Holland, Amsterdam},
  \bibinfo{year}{1985}).

\bibitem[{\citenamefont{Lee and Ramakrishnan}(1985)}]{Lee1985}
\bibinfo{author}{\bibfnamefont{P.~A.} \bibnamefont{Lee}} \bibnamefont{and}
  \bibinfo{author}{\bibfnamefont{T.~V.} \bibnamefont{Ramakrishnan}},
  \bibinfo{journal}{Reviews of Modern Physics} \textbf{\bibinfo{volume}{57}},
  \bibinfo{pages}{287} (\bibinfo{year}{1985}).

\bibitem[{\citenamefont{Potok et~al.}(2006)\citenamefont{Potok, Rau, Shtrikman,
  Oreg, and Goldhaber-Gordon}}]{Potok.06}
\bibinfo{author}{\bibfnamefont{R.~M.} \bibnamefont{Potok}},
  \bibinfo{author}{\bibfnamefont{I.~G.} \bibnamefont{Rau}},
  \bibinfo{author}{\bibfnamefont{H.}~\bibnamefont{Shtrikman}},
  \bibinfo{author}{\bibfnamefont{Y.}~\bibnamefont{Oreg}}, \bibnamefont{and}
  \bibinfo{author}{\bibfnamefont{D.}~\bibnamefont{Goldhaber-Gordon}},
  \bibinfo{journal}{Nature} \textbf{\bibinfo{volume}{446}},
  \bibinfo{pages}{167} (\bibinfo{year}{2006}).

\bibitem[{\citenamefont{Lin and Bird}(2002)}]{Lin2002}
\bibinfo{author}{\bibfnamefont{J.~J.} \bibnamefont{Lin}} \bibnamefont{and}
  \bibinfo{author}{\bibfnamefont{J.~P.} \bibnamefont{Bird}},
  \bibinfo{journal}{Journal of Physics: Condensed Matter}
  \textbf{\bibinfo{volume}{14}}, \bibinfo{pages}{R501} (\bibinfo{year}{2002}).

\bibitem[{\citenamefont{Cichorek et~al.}(2005)\citenamefont{Cichorek, Sanchez,
  Gegenwart, Weickert, Wojakowski, Henkie, Auffermann, Paschen, Kniep, and
  Steglich}}]{Cichorek2005}
\bibinfo{author}{\bibfnamefont{T.}~\bibnamefont{Cichorek}},
  \bibinfo{author}{\bibfnamefont{A.}~\bibnamefont{Sanchez}},
  \bibinfo{author}{\bibfnamefont{P.}~\bibnamefont{Gegenwart}},
  \bibinfo{author}{\bibfnamefont{F.}~\bibnamefont{Weickert}},
  \bibinfo{author}{\bibfnamefont{A.}~\bibnamefont{Wojakowski}},
  \bibinfo{author}{\bibfnamefont{Z.}~\bibnamefont{Henkie}},
  \bibinfo{author}{\bibfnamefont{G.}~\bibnamefont{Auffermann}},
  \bibinfo{author}{\bibfnamefont{S.}~\bibnamefont{Paschen}},
  \bibinfo{author}{\bibfnamefont{R.}~\bibnamefont{Kniep}}, \bibnamefont{and}
  \bibinfo{author}{\bibfnamefont{F.}~\bibnamefont{Steglich}},
  \bibinfo{journal}{Physical Review Letters} \textbf{\bibinfo{volume}{94}},
  \bibinfo{pages}{236603 and references therein} (\bibinfo{year}{2005}).

\bibitem[{\citenamefont{Czulucki et~al.}(2010)\citenamefont{Czulucki,
  Auffermann, Bednarski, Bochenek, B\"{o}hme, Cichorek, Niewa, Oeschler,
  Schmidt, Steglich et~al.}}]{Czulucki2010}
\bibinfo{author}{\bibfnamefont{A.}~\bibnamefont{Czulucki}},
  \bibinfo{author}{\bibfnamefont{G.}~\bibnamefont{Auffermann}},
  \bibinfo{author}{\bibfnamefont{M.}~\bibnamefont{Bednarski}},
  \bibinfo{author}{\bibfnamefont{L.}~\bibnamefont{Bochenek}},
  \bibinfo{author}{\bibfnamefont{M.}~\bibnamefont{B\"{o}hme}},
  \bibinfo{author}{\bibfnamefont{T.}~\bibnamefont{Cichorek}},
  \bibinfo{author}{\bibfnamefont{R.}~\bibnamefont{Niewa}},
  \bibinfo{author}{\bibfnamefont{N.}~\bibnamefont{Oeschler}},
  \bibinfo{author}{\bibfnamefont{M.}~\bibnamefont{Schmidt}},
  \bibinfo{author}{\bibfnamefont{F.}~\bibnamefont{Steglich}},
  \bibnamefont{et~al.}, \bibinfo{journal}{Chemphyschem : a European journal of
  chemical physics and physical chemistry} \textbf{\bibinfo{volume}{11}},
  \bibinfo{pages}{2639 and references therein} (\bibinfo{year}{2010}).

\bibitem[{\citenamefont{Schlechte et~al.}(2007)\citenamefont{Schlechte, Niewa,
  Schmidt, Auffermann, Prots, Schnelle, Gnida, Cichorek, Steglich, and
  Kniep}}]{Schlechte2007}
\bibinfo{author}{\bibfnamefont{A.}~\bibnamefont{Schlechte}},
  \bibinfo{author}{\bibfnamefont{R.}~\bibnamefont{Niewa}},
  \bibinfo{author}{\bibfnamefont{M.}~\bibnamefont{Schmidt}},
  \bibinfo{author}{\bibfnamefont{G.}~\bibnamefont{Auffermann}},
  \bibinfo{author}{\bibfnamefont{Y.}~\bibnamefont{Prots}},
  \bibinfo{author}{\bibfnamefont{W.}~\bibnamefont{Schnelle}},
  \bibinfo{author}{\bibfnamefont{D.}~\bibnamefont{Gnida}},
  \bibinfo{author}{\bibfnamefont{T.}~\bibnamefont{Cichorek}},
  \bibinfo{author}{\bibfnamefont{F.}~\bibnamefont{Steglich}}, \bibnamefont{and}
  \bibinfo{author}{\bibfnamefont{R.}~\bibnamefont{Kniep}},
  \bibinfo{journal}{Science and Technology of Advanced Materials}
  \textbf{\bibinfo{volume}{8}}, \bibinfo{pages}{341} (\bibinfo{year}{2007}).



\bibitem[{\citenamefont{Tremel and Hoffmann}(1987)}]{Tremel1987}
\bibinfo{author}{\bibfnamefont{W.}~\bibnamefont{Tremel}} \bibnamefont{and}
  \bibinfo{author}{\bibfnamefont{R.}~\bibnamefont{Hoffmann}},
  \bibinfo{journal}{Journal of the American Chemical Society}
  \textbf{\bibinfo{volume}{109}}, \bibinfo{pages}{124} (\bibinfo{year}{1987}).

\bibitem[{\citenamefont{Gogolin}(1996)}]{Gogolin.96}
\bibinfo{author}{\bibfnamefont{A.}~\bibnamefont{Gogolin}},
  \bibinfo{journal}{Phys.~Rev.~B} \textbf{\bibinfo{volume}{53}},
  \bibinfo{pages}{R5990} (\bibinfo{year}{1996}).

\bibitem[{\citenamefont{Moustakas and Fisher}(1997)}]{Moustakas.97}
\bibinfo{author}{\bibfnamefont{A.~L.} \bibnamefont{Moustakas}}
  \bibnamefont{and} \bibinfo{author}{\bibfnamefont{D.~S.}
  \bibnamefont{Fisher}}, \bibinfo{journal}{Phys.~Rev.~B}
  \textbf{\bibinfo{volume}{55}}, \bibinfo{pages}{6832} (\bibinfo{year}{1997}).

\bibitem[{\citenamefont{Hotta}(2006)}]{Hotta2006}
\bibinfo{author}{\bibfnamefont{T.}~\bibnamefont{Hotta}},
  \bibinfo{journal}{Physical Review Letters} \textbf{\bibinfo{volume}{96}},
  \bibinfo{pages}{197201} (\bibinfo{year}{2006}).

\bibitem[{\citenamefont{Arnold et~al.}(2007)\citenamefont{Arnold, Langenbruch,
  and Kroha}}]{Arnold2007}
\bibinfo{author}{\bibfnamefont{M.}~\bibnamefont{Arnold}},
  \bibinfo{author}{\bibfnamefont{T.}~\bibnamefont{Langenbruch}},
  \bibnamefont{and} \bibinfo{author}{\bibfnamefont{J.}~\bibnamefont{Kroha}},
  \bibinfo{journal}{Physical Review Letters} \textbf{\bibinfo{volume}{99}},
  \bibinfo{pages}{186601} (\bibinfo{year}{2007}).

\bibitem[{\citenamefont{Chuo et~al.}(2014)\citenamefont{Chuo, Ballmann, Borda,
  and Kroha}}]{FuhChio.14}
\bibinfo{author}{\bibfnamefont{E.~F.} \bibnamefont{Chuo}},
  \bibinfo{author}{\bibfnamefont{K.}~\bibnamefont{Ballmann}},
  \bibinfo{author}{\bibfnamefont{L.}~\bibnamefont{Borda}}, \bibnamefont{and}
  \bibinfo{author}{\bibfnamefont{J.}~\bibnamefont{Kroha}},
  \bibinfo{journal}{J.~Phys.~Conf.~Ser.} \textbf{\bibinfo{volume}{568}},
  \bibinfo{pages}{12007} (\bibinfo{year}{2014}).

\bibitem[{\citenamefont{Sellier et~al.}(2001)\citenamefont{Sellier, Kirchner,
  and Kroha}}]{Sellier2001}
\bibinfo{author}{\bibfnamefont{G.}~\bibnamefont{Sellier}},
  \bibinfo{author}{\bibfnamefont{S.}~\bibnamefont{Kirchner}}, \bibnamefont{and}
  \bibinfo{author}{\bibfnamefont{J.}~\bibnamefont{Kroha}}, in
  \emph{\bibinfo{booktitle}{Kondo Effect and Dephasing in Low-Dimensional
  Metallic Systems}}, edited by
  \bibinfo{editor}{\bibfnamefont{V.}~\bibnamefont{Chandrasekhar}},
  \bibinfo{editor}{\bibfnamefont{C.~V.} \bibnamefont{Haesendonck}},
  \bibnamefont{and}
  \bibinfo{editor}{\bibfnamefont{A.}~\bibnamefont{Zawadowski}}
  (\bibinfo{publisher}{Springer Netherlands}, \bibinfo{year}{2001}), pp.
  \bibinfo{pages}{241--244}.

\bibitem[{Sup()}]{SupMat}
\bibinfo{note}{See Supplemental Material.}

%
%


















%
\end{thebibliography}

\begin{thebibliography}{0}
\expandafter\ifx\csname natexlab\endcsname\relax\def\natexlab#1{#1}\fi
\expandafter\ifx\csname bibnamefont\endcsname\relax
  \def\bibnamefont#1{#1}\fi
\expandafter\ifx\csname bibfnamefont\endcsname\relax
  \def\bibfnamefont#1{#1}\fi
\expandafter\ifx\csname citenamefont\endcsname\relax
  \def\citenamefont#1{#1}\fi
\expandafter\ifx\csname url\endcsname\relax
  \def\url#1{\texttt{#1}}\fi
\expandafter\ifx\csname urlprefix\endcsname\relax\def\urlprefix{URL }\fi
\providecommand{\bibinfo}[2]{#2}
\providecommand{\eprint}[2][]{\url{#2}}

\bibitem[{\citenamefont{Binnewies et~al.}(2012)\citenamefont{Binnewies, Glaum,
  Schmidt, and Schmidt}}]{Binnewies}
\bibinfo{author}{\bibfnamefont{M.}~\bibnamefont{Binnewies}},
  \bibinfo{author}{\bibfnamefont{R.}~\bibnamefont{Glaum}},
  \bibinfo{author}{\bibfnamefont{M.}~\bibnamefont{Schmidt}}, \bibnamefont{and}
  \bibinfo{author}{\bibfnamefont{P.}~\bibnamefont{Schmidt}},
  \emph{\bibinfo{title}{Chemical Vapor Transport Reactions}}
  (\bibinfo{publisher}{Walter de Gruyter}, \bibinfo{year}{2012}).

\bibitem[{\citenamefont{Schlechte et~al.}(2007)\citenamefont{Schlechte, Niewa,
  Schmidt, Auffermann, Prots, Schnelle, Gnida, Cichorek, Steglich, and
  Kniep}}]{Schlechte2007}
\bibinfo{author}{\bibfnamefont{A.}~\bibnamefont{Schlechte}},
  \bibinfo{author}{\bibfnamefont{R.}~\bibnamefont{Niewa}},
  \bibinfo{author}{\bibfnamefont{M.}~\bibnamefont{Schmidt}},
  \bibinfo{author}{\bibfnamefont{G.}~\bibnamefont{Auffermann}},
  \bibinfo{author}{\bibfnamefont{Y.}~\bibnamefont{Prots}},
  \bibinfo{author}{\bibfnamefont{W.}~\bibnamefont{Schnelle}},
  \bibinfo{author}{\bibfnamefont{D.}~\bibnamefont{Gnida}},
  \bibinfo{author}{\bibfnamefont{T.}~\bibnamefont{Cichorek}},
  \bibinfo{author}{\bibfnamefont{F.}~\bibnamefont{Steglich}}, \bibnamefont{and}
  \bibinfo{author}{\bibfnamefont{R.}~\bibnamefont{Kniep}},
  \bibinfo{journal}{Science and Technology of Advanced Materials}
  \textbf{\bibinfo{volume}{8}}, \bibinfo{pages}{341} (\bibinfo{year}{2007}),
  ISSN \bibinfo{issn}{1468-6996},

\bibitem[{\citenamefont{Wang and Hughbanks}(1995)}]{Wang.95}
\bibinfo{author}{\bibfnamefont{C.}~\bibnamefont{Wang}} \bibnamefont{and}
  \bibinfo{author}{\bibfnamefont{T.}~\bibnamefont{Hughbanks}},
  \bibinfo{journal}{Inorg.~Chem.} \textbf{\bibinfo{volume}{34}},
  \bibinfo{pages}{5524} (\bibinfo{year}{1995}).

\bibitem[{\citenamefont{Schlechte et~al.}(2009)\citenamefont{Schlechte, Niewa,
  Prots, Schnelle, Schmidt, and Kniep}}]{Schlechte2009}
\bibinfo{author}{\bibfnamefont{A.}~\bibnamefont{Schlechte}},
  \bibinfo{author}{\bibfnamefont{R.}~\bibnamefont{Niewa}},
  \bibinfo{author}{\bibfnamefont{Y.}~\bibnamefont{Prots}},
  \bibinfo{author}{\bibfnamefont{W.}~\bibnamefont{Schnelle}},
  \bibinfo{author}{\bibfnamefont{M.}~\bibnamefont{Schmidt}}, \bibnamefont{and}
  \bibinfo{author}{\bibfnamefont{R.}~\bibnamefont{Kniep}},
  \bibinfo{journal}{Inorganic Chemistry} \textbf{\bibinfo{volume}{48}},
  \bibinfo{pages}{2277} (\bibinfo{year}{2009}), ISSN \bibinfo{issn}{1520-510X},

\bibitem[{\citenamefont{Tremel and Hoffmann}(1987)}]{Tremel1987}
\bibinfo{author}{\bibfnamefont{W.}~\bibnamefont{Tremel}} \bibnamefont{and}
  \bibinfo{author}{\bibfnamefont{R.}~\bibnamefont{Hoffmann}},
  \bibinfo{journal}{Journal of the American Chemical Society}
  \textbf{\bibinfo{volume}{109}}, \bibinfo{pages}{124} (\bibinfo{year}{1987}),
  ISSN \bibinfo{issn}{0002-7863},

\bibitem[{\citenamefont{Doert et~al.}(2012)\citenamefont{Doert, Graf,
  Vasilyeva, and Schnelle}}]{Doert2012}
\bibinfo{author}{\bibfnamefont{T.}~\bibnamefont{Doert}},
  \bibinfo{author}{\bibfnamefont{C.}~\bibnamefont{Graf}},
  \bibinfo{author}{\bibfnamefont{I.~G.} \bibnamefont{Vasilyeva}},
  \bibnamefont{and} \bibinfo{author}{\bibfnamefont{W.}~\bibnamefont{Schnelle}},
  \bibinfo{journal}{Inorganic Chemistry} \textbf{\bibinfo{volume}{51}},
  \bibinfo{pages}{282} (\bibinfo{year}{2012}), ISSN \bibinfo{issn}{1520-510X},

\bibitem[{\citenamefont{Akkermans and Montambaux}(2006)}]{Akkermans}
\bibinfo{author}{\bibfnamefont{E.}~\bibnamefont{Akkermans}} \bibnamefont{and}
  \bibinfo{author}{\bibfnamefont{G.}~\bibnamefont{Montambaux}},
  \emph{\bibinfo{title}{Mesoscopic Physics of electrons and photons}}
  (\bibinfo{publisher}{Cambridge University Press}, \bibinfo{year}{2006}).

\bibitem[{\citenamefont{Reim}(1986)}]{Reim.86}
\bibinfo{author}{\bibfnamefont{W.}~\bibnamefont{Reim}},
  \bibinfo{journal}{J.~Magn.~Magn.~Mater.} \textbf{\bibinfo{volume}{58}},
  \bibinfo{pages}{1} (\bibinfo{year}{1986}).

\bibitem[{\citenamefont{Cichorek et~al.}(2002)\citenamefont{Cichorek, Henkie,
  Wojakowski, Pietraszko, Gegenwart, Lang, and Steglich}}]{Cichorek2002}
\bibinfo{author}{\bibfnamefont{T.}~\bibnamefont{Cichorek}},
  \bibinfo{author}{\bibfnamefont{Z.}~\bibnamefont{Henkie}},
  \bibinfo{author}{\bibfnamefont{A.}~\bibnamefont{Wojakowski}},
  \bibinfo{author}{\bibfnamefont{A.}~\bibnamefont{Pietraszko}},
  \bibinfo{author}{\bibfnamefont{P.}~\bibnamefont{Gegenwart}},
  \bibinfo{author}{\bibfnamefont{M.}~\bibnamefont{Lang}}, \bibnamefont{and}
  \bibinfo{author}{\bibfnamefont{F.}~\bibnamefont{Steglich}},
  \bibinfo{journal}{Solid State Communications} \textbf{\bibinfo{volume}{121}},
  \bibinfo{pages}{647} (\bibinfo{year}{2002}), ISSN \bibinfo{issn}{00381098},

\bibitem[{\citenamefont{Altshuler and Aronov}(1985)}]{Altshuler1985}
\bibinfo{author}{\bibfnamefont{B.}~\bibnamefont{Altshuler}} \bibnamefont{and}
  \bibinfo{author}{\bibfnamefont{A.}~\bibnamefont{Aronov}},
  \emph{\bibinfo{title}{{Electron-Electron Interactions in Disordered
  Systems}}} (\bibinfo{publisher}{North-Holland, Amsterdam},
  \bibinfo{year}{1985}).

\bibitem[{\citenamefont{Lee and Ramakrishnan}(1985)}]{Lee1985}
\bibinfo{author}{\bibfnamefont{P.~A.} \bibnamefont{Lee}} \bibnamefont{and}
  \bibinfo{author}{\bibfnamefont{T.~V.} \bibnamefont{Ramakrishnan}},
  \bibinfo{journal}{Reviews of Modern Physics} \textbf{\bibinfo{volume}{57}},
  \bibinfo{pages}{287} (\bibinfo{year}{1985}), ISSN \bibinfo{issn}{0034-6861},

\bibitem[{\citenamefont{Lin and Bird}(2002)}]{Lin2002}
\bibinfo{author}{\bibfnamefont{J.~J.} \bibnamefont{Lin}} \bibnamefont{and}
  \bibinfo{author}{\bibfnamefont{J.~P.} \bibnamefont{Bird}},
  \bibinfo{journal}{Journal of Physics: Condensed Matter}
  \textbf{\bibinfo{volume}{14}}, \bibinfo{pages}{R501} (\bibinfo{year}{2002}),
  ISSN \bibinfo{issn}{09538984},

\bibitem[{\citenamefont{Micklitz et~al.}(2006)\citenamefont{Micklitz, Altland,
  Costi, and Rosch}}]{Micklitz2006}
\bibinfo{author}{\bibfnamefont{T.}~\bibnamefont{Micklitz}},
  \bibinfo{author}{\bibfnamefont{A.}~\bibnamefont{Altland}},
  \bibinfo{author}{\bibfnamefont{T.}~\bibnamefont{Costi}}, \bibnamefont{and}
  \bibinfo{author}{\bibfnamefont{A.}~\bibnamefont{Rosch}},
  \bibinfo{journal}{Physical Review Letters} \textbf{\bibinfo{volume}{96}},
  \bibinfo{pages}{226601} (\bibinfo{year}{2006}), ISSN
  \bibinfo{issn}{0031-9007},

\bibitem[{\citenamefont{Zawadowski et~al.}(1999)\citenamefont{Zawadowski, von
  Delft, and Ralph}}]{Zawadowski1999}
\bibinfo{author}{\bibfnamefont{A.}~\bibnamefont{Zawadowski}},
  \bibinfo{author}{\bibfnamefont{J.}~\bibnamefont{von Delft}},
  \bibnamefont{and} \bibinfo{author}{\bibfnamefont{D.}~\bibnamefont{Ralph}},
  \bibinfo{journal}{Physical Review Letters} \textbf{\bibinfo{volume}{83}},
  \bibinfo{pages}{2632} (\bibinfo{year}{1999}), ISSN \bibinfo{issn}{0031-9007},

\bibitem[{\citenamefont{Schmidt et~al.}(2005)\citenamefont{Schmidt, Cichorek,
  Niewa, Schlechte, Prots, Steglich, and Kniep}}]{Schmidt2005a}
\bibinfo{author}{\bibfnamefont{M.}~\bibnamefont{Schmidt}},
  \bibinfo{author}{\bibfnamefont{T.}~\bibnamefont{Cichorek}},
  \bibinfo{author}{\bibfnamefont{R.}~\bibnamefont{Niewa}},
  \bibinfo{author}{\bibfnamefont{A.}~\bibnamefont{Schlechte}},
  \bibinfo{author}{\bibfnamefont{Y.}~\bibnamefont{Prots}},
  \bibinfo{author}{\bibfnamefont{F.}~\bibnamefont{Steglich}}, \bibnamefont{and}
  \bibinfo{author}{\bibfnamefont{R.}~\bibnamefont{Kniep}},
  \bibinfo{journal}{Journal of Physics: Condensed Matter}
  \textbf{\bibinfo{volume}{17}}, \bibinfo{pages}{5481} (\bibinfo{year}{2005}),
  ISSN \bibinfo{issn}{0953-8984},

\bibitem[{\citenamefont{Gogolin}(1996)}]{Gogolin.96}
\bibinfo{author}{\bibfnamefont{A.}~\bibnamefont{Gogolin}},
  \bibinfo{journal}{Phys.~Rev.~B} \textbf{\bibinfo{volume}{53}},
  \bibinfo{pages}{R5990} (\bibinfo{year}{1996}).

\bibitem[{\citenamefont{Hotta}(2006)}]{Hotta2006}
\bibinfo{author}{\bibfnamefont{T.}~\bibnamefont{Hotta}},
  \bibinfo{journal}{Physical Review Letters} \textbf{\bibinfo{volume}{96}},
  \bibinfo{pages}{197201} (\bibinfo{year}{2006}), ISSN
  \bibinfo{issn}{0031-9007},

\bibitem[{\citenamefont{Vlad\'{a}r and
  Zawadowski}(1983{\natexlab{a}})}]{Vladar1983a}
\bibinfo{author}{\bibfnamefont{K.}~\bibnamefont{Vlad\'{a}r}} \bibnamefont{and}
  \bibinfo{author}{\bibfnamefont{A.}~\bibnamefont{Zawadowski}},
  \bibinfo{journal}{Physical Review B} \textbf{\bibinfo{volume}{28}},
  \bibinfo{pages}{1596} (\bibinfo{year}{1983}{\natexlab{a}}), ISSN
  \bibinfo{issn}{0163-1829},

\bibitem[{\citenamefont{Vlad\'{a}r and
  Zawadowski}(1983{\natexlab{b}})}]{Vladar1983b}
\bibinfo{author}{\bibfnamefont{K.}~\bibnamefont{Vlad\'{a}r}} \bibnamefont{and}
  \bibinfo{author}{\bibfnamefont{A.}~\bibnamefont{Zawadowski}},
  \bibinfo{journal}{Physical Review B} \textbf{\bibinfo{volume}{28}},
  \bibinfo{pages}{1564} (\bibinfo{year}{1983}{\natexlab{b}}), ISSN
  \bibinfo{issn}{0163-1829},

\bibitem[{\citenamefont{Zawadowski}(1980)}]{Zawadowski1980}
\bibinfo{author}{\bibfnamefont{A.}~\bibnamefont{Zawadowski}},
  \bibinfo{journal}{Physical Review Letters} \textbf{\bibinfo{volume}{45}},
  \bibinfo{pages}{211} (\bibinfo{year}{1980}), ISSN \bibinfo{issn}{0031-9007},

\bibitem[{\citenamefont{Cox and Zawadowski}(1998)}]{Cox1998}
\bibinfo{author}{\bibfnamefont{D.~L.} \bibnamefont{Cox}} \bibnamefont{and}
  \bibinfo{author}{\bibfnamefont{A.}~\bibnamefont{Zawadowski}},
  \bibinfo{journal}{Advances in Physics} \textbf{\bibinfo{volume}{47}},
  \bibinfo{pages}{599} (\bibinfo{year}{1998}), ISSN \bibinfo{issn}{0001-8732},

\bibitem[{\citenamefont{Aleiner and Controzzi}(2002)}]{Aleiner2002}
\bibinfo{author}{\bibfnamefont{I.}~\bibnamefont{Aleiner}} \bibnamefont{and}
  \bibinfo{author}{\bibfnamefont{D.}~\bibnamefont{Controzzi}},
  \bibinfo{journal}{Physical Review B} \textbf{\bibinfo{volume}{66}},
  \bibinfo{pages}{045107} (\bibinfo{year}{2002}), ISSN
  \bibinfo{issn}{0163-1829},

\bibitem[{\citenamefont{Moustakas and Fisher}(1997)}]{Moustakas.97}
\bibinfo{author}{\bibfnamefont{A.~L.} \bibnamefont{Moustakas}}
  \bibnamefont{and} \bibinfo{author}{\bibfnamefont{D.~S.}
  \bibnamefont{Fisher}}, \bibinfo{journal}{Phys.~Rev.~B}
  \textbf{\bibinfo{volume}{55}}, \bibinfo{pages}{6832} (\bibinfo{year}{1997}).

\bibitem[{\citenamefont{Arnold et~al.}(2007)\citenamefont{Arnold, Langenbruch,
  and Kroha}}]{Arnold2007}
\bibinfo{author}{\bibfnamefont{M.}~\bibnamefont{Arnold}},
  \bibinfo{author}{\bibfnamefont{T.}~\bibnamefont{Langenbruch}},
  \bibnamefont{and} \bibinfo{author}{\bibfnamefont{J.}~\bibnamefont{Kroha}},
  \bibinfo{journal}{Physical Review Letters} \textbf{\bibinfo{volume}{99}},
  \bibinfo{pages}{186601} (\bibinfo{year}{2007}), ISSN
  \bibinfo{issn}{0031-9007},

\bibitem[{\citenamefont{Abrikosov and Gor'kov}(1961)}]{Abrikosov1961}
\bibinfo{author}{\bibfnamefont{A.~A.} \bibnamefont{Abrikosov}}
  \bibnamefont{and} \bibinfo{author}{\bibfnamefont{L.~P.}
  \bibnamefont{Gor'kov}}, \bibinfo{journal}{Sov. Phys.-JETP}
  \textbf{\bibinfo{volume}{12}}, \bibinfo{pages}{1243} (\bibinfo{year}{1961}).

\bibitem[{\citenamefont{Sellier et~al.}(2001)\citenamefont{Sellier, Kirchner,
  and Kroha}}]{Sellier2001}
\bibinfo{author}{\bibfnamefont{G.}~\bibnamefont{Sellier}},
  \bibinfo{author}{\bibfnamefont{S.}~\bibnamefont{Kirchner}}, \bibnamefont{and}
  \bibinfo{author}{\bibfnamefont{J.}~\bibnamefont{Kroha}}, in
  \emph{\bibinfo{booktitle}{Kondo Effect and Dephasing in Low-Dimensional
  Metallic Systems}}, edited by
  \bibinfo{editor}{\bibfnamefont{V.}~\bibnamefont{Chandrasekhar}},
  \bibinfo{editor}{\bibfnamefont{C.~V.} \bibnamefont{Haesendonck}},
  \bibnamefont{and}
  \bibinfo{editor}{\bibfnamefont{A.}~\bibnamefont{Zawadowski}}
  (\bibinfo{publisher}{Springer Netherlands}, \bibinfo{year}{2001}), pp.
  \bibinfo{pages}{241--244},

\end{thebibliography}
\end{document}